\documentclass[preprintnumbers,amsmath,amssymb,subfloat,floatfix,10pt,prd,superscriptaddress,nofootinbib]{revtex4}
\usepackage{graphicx,epsfig} 
\usepackage{dcolumn}
\usepackage{bm}
\usepackage{caption}
\usepackage{subfig}
\usepackage{amssymb}
\usepackage{color}
\usepackage{amsmath}

\begin{document}


\title{ Effect of low anisotropy on cosmological models  by using supernova data  }
\author{H. Hossienkhani}
\email{hossienhossienkhani@yahoo.com}
\affiliation{Department of Physics, Hamedan Branch, Islamic Azad University, Hamedan, Iran}
\author{H. Yousefi}
\email{h.yousefi@hut.ac.ir}
\affiliation{Department of Science, Hamedan University of Technology,  Hamedan, 65155, Iran}
\author{N. Azimi}
\email{azimi1379@yahoo.com}
\affiliation{Department of Mathematics, Hamedan Branch, Islamic Azad University, Hamedan, Iran}

\date{\today}

\begin{abstract}
\vspace*{1.5cm} \centerline{\bf Abstract} \vspace*{.5cm}

\end{abstract}


\maketitle
By using the supernovae type Ia data we study    influence of the anisotropy (although low) on the evolution of the universe and   compare $\Lambda$CDM model with 6 representative parametrizations of the recent Hubble expansion history $H(z)$.  To compare these models we use  the  the maximum likelihood  method  for find that the best fit dynamical  $w(z)$ and $q(z)$ obtained from the SNIa dataset.  We have performed a comparative analysis of two SNIa datasets such as, the 194 SNIa
($0\leq z\leq1.75$) and the most recent Supernova Legacy Survey (SNLS) dataset (238 data points  $0.15< z<1.1$). In particular we find the best fit  value of $\Lambda$CDM model $\Omega_{\sigma_0}=0.013$,~ $\chi^2_{min}=197.56$ with 194 SNIa   and  $\Omega_{\sigma_0}=-0.003\pm0.033$,~ $\chi^2_{min}=230.656$   with 238 SNLS high redshift type Ia Supernovae.  The analysis shows that   by considering the anisotropy, it  leads to more  best fit parameters  in all   models (except of SCDM)  with  SNIa data.
We also use two statistical tests such as the usual  $\chi^2_{min}/dof$   and p-test to compare different dark energy models. According to
both statistical tests and considering anisotropy,  the $\Lambda$CDM   model  that is  providing the best fit to  $\Lambda$CDM of FRW model. An even better fit
would result with an optimization of the data using effects of anisotropy from the beginning.
\\

{\bf{keywords}:}{  Anisotropic universe,  Dark energy,  Supernovae type Ia, Luminosity distance}


\section{Introductions}
It is well known that  the universe is currently undergoing accelerated cosmological expansion. Scientific observation shows that the accelerated expansion comes from supernova Type Ia (SNIa) apparent magnitude measurements as
a function of redshift \cite{1,2,3}, cosmic microwave background (CMB) anisotropy data \cite{4,5,6} combined with low estimates
of the cosmological mass density \cite{7}, and baryon acoustic oscillation (BAO) peak length scale estimates \cite{8,9}.
Based on the years of analyzing the first year Wilkinson Microwave Anisotropy Probe (WMAP) data the observational results are formed  \cite{10,11}.
By the introduction of a nonvanishing cosmological constant,  the accelerated expansion is taken into account $\Lambda$, where observations suggest a value of
 $\Lambda=10^{-52}m^{-2}$ \cite{12}  in standard cosmology. The role of this cosmological constant is a non-clustering energy form, generally referred to dark energy (DE). According to the observations and standard particle physics theory, standard cosmology, including a universe filled by a cosmological constant along with the cold dark matter (CDM) and baryonic matters   \cite{13}.  The simplest extension to $\Lambda$ is the DE with a constant $w$, of which the corresponding cosmological model is the so-called $w$CDM model, and a slowly rolling DE scalar field with an inverse power-law potential ($\phi$CDM model) \cite{14a,14}.   Also a scalar field $\phi$ plays the role of DE, in the $\phi$CDM scenario, as far as while spatial curvature is allowed to be non-zero in the  $\Lambda$CDM case, we consider a spatially-flat cosmological model, in the $w$CDM and $\phi$CDM cases. Different types of DE models have been reviewed previously such as quintessence \cite{14,15,16,17,18,19,19a}, phantom \cite{ 20,21,22}, K-essence \cite{23,24,25} and etc. In addition, the Chaplygin gas (CG) \cite{26,27,28} and generalized Chaplygin gas (GCG) are considered extensively as an interaction between DE and dark matter (DM)   \cite{29,30,31,32}. So far, from observations of high redshift SNIa, there is a tremendous amount of activity going on in trying to determine  the equation of state $w_X(z)$ and other cosmological parameters \cite{33,34,35,36,37,38,38a}. The most important way to measure the history  of the cosmic expansion, is measurement of  the luminocity distance  relation. Since  SNIa data measures the luminosity distance redshift relationship, it
provides a purely kinematic record of the expansion history of the universe.  By using SNIa data without assuming the nature and evolution of the DE, probing the evolution of the Hubble parameter or the deceleration parameter is possible  \cite{39,40}.  These result are based on fitting a Friedman-Robertson-Walker (FRW) type geometry, together with the corresponding cosmology, to the existing astronomical data. \\
It is natural to assume that the geometry at very early epoch more general than just the isotropic and homogeneous FRW. Although on large scale at present, the universe seems homogeneous and isotropic, to guarantee of the isotropy there is no observational data in an era prior to the recombination.  In fact, it is possible to begin with an
anisotropic universe which isotropizes during its evolution. Jaffe  \textit{et al}. \cite{41} research's show that removing a Bianchi component from the WMAP data can account for several large-angle anomalies leaving the universe to be isotropic. Therefore regardless of the inflation, in cosmological models, the universe may have achieved a slight anisotropic geometry.  The Bianchi universe models \cite{42} are spatially homogeneous anisotropic cosmological models.
From observational data, several strong limits on anisotropic models are investigated \cite{43,44,44a}. Sharif and Saleem \cite{44b} studied warm inflation for the Bianchi type I (BI) model and showed that this model is consistent with observational data.  Also, Sharif and  Siddiqa \cite{44c}  examined
 effects of viscosity on anisotropic universe in modified gravity and observed  that bulk viscosity enhances expansion of the universe. 
 Recently, Hossienkhani \textit{et al}. \cite{45} discussed the effects of the
  anisotropy on the evolutionary behavior DE models and compare with the results of the standard FRW, $\Lambda$CDM and $w$CDM
models. Also, they shown that the    anisotropy is a non-zero value at the present time $a=1$ although it is approaching zero, i.e. the anisotropy
will be very low after inflation. Hence, the effects of anisotropy can be investigated in the context of DE and DM models although it is low. In this paper  the usefulness of anisotropy effect to assess the parameters of several popular DE models investigate
and compare our results with the inference made
using luminosity distances measured with SNIa.\\
The paper is organized as follows. In the next section, we present the    field equations for  the BI universe and derive the cosmological evolution of
  the equation of state parameter $w(z)$ and  the deceleration parameter $q(z)$ for several DE models. In section III we fit the derived Hubble parameter to the
SNIa   dataset  and obtain constraints for the various DE models in a flat BI model.
In section IV we present our results for the best fits and discuss their implications and common features.  This analysis is repeated with data obtained from
observationally allowed thawing models even though this class of models has small allowed deviations from $\Lambda$CDM model.
 In section V  we briefly summarize the contents of the SNLS sample and describe  the light-curve fitting method and the model parameters (including those associated with the data) that are to be estimated. The last section is devoted to summary and conclusions.

\section{Anisotropy effects  and cosmological analysis using   SNIa data}
In our analysis, the most typical and commonly used current observations are chosen, i.e., the type Ia supernovae (SNIa), and direct measurement of the luminocity distance.
In other words, SNIa can be calibrated to be good cosmological standard candles, with small dispersions in their peak luminocity \cite{46}. We begin with a brief outline of the method of our analysis of the supernova data. The luminosity distance $d_L(z)$ to an object at redshift $z$ is such that
\begin{equation}\label{1}
d_L(z)=r(z)(1+z),
\end{equation}
where $r(z)$ is the comoving distance and  the relation between the light ray geodesic and  the comoving distance in in a flat universe  is $cdt=a(z)dr(z)$, where $a$ is the scale factor. The apparent magnitude $m(z)$ of the source with an absolute magnitude $M$ is related to the luminosity distance $d_L(z)$  with  \cite{47,48}
\begin{equation}\label{2}
\mu_{th}\equiv m(z)-M=5log_{10}\bigg(\frac{d_L(z)}{Mpc}\bigg)+25.
\end{equation}
The  absolute magnitude  $M$   after correcting for supernova light curve width $\textbf{-}$ luminosity correlation \cite{2,49}. After applying the
above correction, $M$, and hence $m(z)$, is believed to be constant for all SNIa. In the case of SNIa, the first step is almost trivial since the textbook expression
for $D_L(z)$ reads
\begin{equation}\label{3}
D_L(z)=\frac{H_0 d_L(z)}{c}=(1+z)\int^z_0 dz'\frac{H_0}{H(z')},
\end{equation}
where $H(z)=H(z;\theta)$ is the Hubble parameter as a function of redshift and  terms of arbitrary parameters. The  predicted $d_L(z)$  can be compared with the observed $d_L(z)$  to test the consistency of the theoretical model with observations. In this paper,
we would like to use both the maximum likelihood estimation (MLE)  \cite{50}  and the   SNIa of the   dataset  compiled in  \cite{47}. So, we can find
the goodness of fit to the corresponding observed $D_L(z_i)$ ($i = 1, ..., 194$) coming from the SNIa data  \cite{47}.
The goodness of fit corresponding to any slope $\theta$ is determined by the probability distribution of $\theta$ i.e. \cite{50}
\begin{equation}\label{4}
P(\theta)=\mathcal{N}e^{-\frac{1}{2}\chi^2(\theta)},
\end{equation}
where $\mathcal{N}$ is a normalization constant.  We also call this measure ``p-test" and it will be used in what follows to compare the quality of the parametrizations considered.
The theoretical model parameters are determined by minimizing the quantity \cite{40a}
\begin{equation}\label{5}
\chi^2(\bar{\theta}')=\sum^N_{i=1}\frac{[log_{10}D_L^{obs}(z_i)-0.2\bar{\theta}'-log_{10}D_L^{th}(z_i)]^2}{[\sigma'_{log_{10}D_L(z_i)}]^2+[\frac{\partial log_{10}D_L(z_i)}{\partial z_i}\sigma'_{z_i}]^2},
\end{equation}
where $N=194$ for SNIa,  $\bar{\theta}'=\theta-\theta_{obs}$ is a free parameter representing
the difference between the actual $\theta$, $\sigma'_z$ and $\sigma'_{log_{10}D_L(z_i)}$ are the $1\sigma'$ redshift uncertainty and errors  of the data and $log_{10}D_L^{obs}(z_i)$ respectively. These errors are assumed to be Gaussian and uncorrelated.  In the case of $\bar{\theta}'$, we assumed no prior constraint on
$\bar{\theta}'$, which is just an unknown constant with a value between $-\infty$ and $+\infty$. In this case, we integrated the probabilities on $\bar{\theta}'$ and therefore worked with a $\bar{\chi}^2$ defined by
\begin{equation}\label{6}
\bar{\chi}^2=-2\ln(\int^{+\infty}_{-\infty}e^{\frac{-\chi^2}{2}}d\bar{\theta}')=A'+\frac{B'^2}{C'}+\ln\frac{C'}{2\pi},
\end{equation}
where
\begin{eqnarray}
&&A'=\sum^{194}_{i=1}\frac{s^2_i}{\sigma'^2_i}=\chi^2(\bar{\theta}'=0),   \label{6a} \\
&&B'=0.2\sum^{194}_{i=1}\frac{s_i}{\sigma'^2_i}, \label{6b}\\
&&C'=0.04\sum^{194}_{i=1}\frac{1}{\sigma'^2_i}, \label{6c}
\end{eqnarray}
with $s_i=log_{10}D_L^{obs}-log_{10}D_L^{th}$. The steps we followed for the usual minimization of Eq. (\ref{5}) in terms of its parameters are
described in detail in \cite{50a,51,52}. This approach assumes that there is some theoretical model available, given in the form of $H(z; \theta_i)$, which is to be compared against the data.
As a result of the analysis, the best-fit parameter values and the corresponding  $1\sigma'$ and  $2\sigma'$  error bars are
obtained. The $1\sigma'$ error on $\theta$ is determined by the relation \cite{50}
\begin{equation}\label{7}
\Delta \chi^2_{1\sigma'}=\chi^2(\theta_{1\sigma'})-\chi^2_{min}=1,
\end{equation}
where the best fit value of $\theta (\theta=\theta_0)$ is given by the value that minimizes $\chi^2(\theta)(\chi^2(\theta_{0})=\chi^2_{min})$. From Eq. (\ref{7}) $\theta$ is in the range $[\theta_0, \theta_{1\sigma'}]$ with 68\% probability for $n=1$, where $n$   is the number of free model parameters.  Also the $2\sigma'$ error  with 95.4\% range which that is determined by
$\Delta \chi^2_{2\sigma'}=4$. \\
To evaluate the influence of both the global expansion and the line of sight conditions on light propagation we
examine an anisotropic accurate solution of the Einstein field equations. The BI cosmology has different expansion rates along the three orthogonal spatial directions, given by the metric
\begin{equation}\label{8}
ds^2=dt^{2}-A^{2}(t)dx^{2}-B^{2}(t)dy^{2}-C^{2}(t)dz^{2},
\end{equation}
where $A(t)$, $B(t)$ and $C(t)$ are the scale factors which describe the anisotropy of the model and the average expansion scale factor $a(t)=(ABC)^{1/3}$. It reduces to the FRW case when $A(t)=B(t)=C(t)=a(t)$.
Defining the time-like hypersurface-orthogonal vector $u=\partial/ \partial t$, we can define the  average Hubble scalar, $H$, and the shear, $\sigma_{\mu\nu}$, as follows:
\begin{eqnarray}\label{9}
~~~~~~H=\frac{1}{3}u_{;\mu}^\mu, \quad~~\sigma_{\mu \nu}=u_{(\mu;\nu)}-H\delta_{\mu\nu}.
\end{eqnarray}
Einstein's field equations  for the  BI metric is given in  (\ref{8}) which lead to the following system of equations \cite{45,53}
\begin{eqnarray}
&&3H^{2}-\sigma^{2}=\kappa^2(\rho_{m}+\rho_{D}),   \label{10} \\
&&3H^2+2\dot{H}+\sigma^{2}=-\kappa^2\left(p_{m}+p_{D}\right), \label{11}\\
&&\dot{\sigma}+3H\sigma=0, \label{12}
\end{eqnarray}
where   $\rho_{D}$ and $p_{D}$ are  the energy density and pressure of DE, respectively. We  also define  $\kappa^2=8\pi G$, here $G$ is Newton's gravitational constant
and will use the unit $\kappa^2=1$.  One can rewrite Eq. (\ref{10}) in the form
\begin{equation}\label{13}
\Omega_m+\Omega_{D}=1-\Omega_{\sigma},
\end{equation}
where
\begin{eqnarray}\label{14}
\Omega_m &=&\frac{8\pi G\rho_m}{3 H^2}=\Omega_{m_0}(1+z)^3(\frac{H_0}{H})^2,\\
\Omega_{\sigma}&=&\frac{\sigma^2}{3H^2}=\Omega_{\sigma_0}(1+z)^6(\frac{H_0}{H})^2,\label{15}
\end{eqnarray}
where $z$ is the redshift, $z = 1/a-1$,  $\Omega_{\sigma_0}$ is
probably the current fractional density due to so-called anisotropy
and $\Omega_{m_0}$  is the current fractional density of non-relativistic
matter.
Eq. (\ref{13}) shows that the sum of the energy density parameters approaches $1$ at late times if the shear
tensor tend zero. Hence, at the late times the universe becomes flat, i.e. for sufficiently large time, this model
predicts that the anisotropy of the universe will damp out and universe will become isotropic. \\
The DE is usually described by an equation of state parameter (EoS) $w(z)=p_D(z)/\rho_D(z)$. Using Eqs. (\ref{10}) and (\ref{11}) we can obtain the expression for EoS parameter
\begin{equation}\label{16}
w(z)=\frac{\frac{2}{3}(1+z)\frac{d\ln H}{dz}-\Omega_{\sigma_0}(1+z)^6(\frac{H_0}{H})^2-1}{1-(\frac{H_0}{H})^2\bigg(\Omega_{m_0}(1+z)^3+\Omega_{\sigma_0}(1+z)^6\bigg)}.
\end{equation}
Independently of its physical origin, the parameter $w(z)$ is an observable derived from $H(z)$  and is usually used to compare theoretical model
predictions with observations. When the anisotropy density goes to zero, i.e. $\sigma_0\rightarrow 0$, and $\Omega_{\sigma_0}\rightarrow 0$ (i.e. spatially
flat FRW  universe), the EoS parameter is reduced to that of the \cite{40a,51,54}. Also, the acceleration of
the universe can be quantified through a   cosmological function known as the deceleration parameter $q$, equivalently
\begin{equation}\label{17}
q(z)=-1+(1+z)\frac{d\ln H}{dz}=\frac{1}{2}+\frac{3}{2}w(z)\bigg[1-(\frac{H_0}{H})^2\bigg(\Omega_{m_0}(1+z)^3+\Omega_{\sigma_0}(1+z)^6\bigg)\bigg]+\frac{3}{2}\Omega_{\sigma_0}(1+z)^6(\frac{H_0}{H})^2,
\end{equation}
where $q<0$ describes an accelerating universe, whereas $q>0$ for a universe which is   decelerating phase.

\section{Likelihood analysis    and   comparison of DE models using   SNIa dataset}
We will consider five representative $H(z)$ parametrizations and minimize the $\chi^2$ of Eqs. (\ref{5}) with respect to
model parameters. We compare the best fit parametrizations obtained with SNIa dataset.  Besides a measure of the quality of fit may be defined in analogy
to the likelihood for   comparison of  DE models to be used with a given set of $w(z)$ data. Refs. \cite{55,56}   motivated by the high likelihood of the phantom divide
line  crossing indicated by the Gold datasets \cite{38,57} to examine theoretical models that predict such crossing. Note that an additional uncertainty from the redshift
dispersion due to peculiar velocity must be added to the uncertainty of each SNIa data point.  So,  one must propagate
$\sigma'_z=c^{-1}500km/sec$ into an additional uncertainty in the luminosity distance. \\
Let us start our analysis with the flat BI model where $\Omega_m+\Omega_{D}+\Omega_{\sigma}=1$, with two free parameters $(\Omega_{\sigma},s_i)$. We have also chosen the parameter $\Omega_{m_0}$ from recent Planck data observed $\Omega_{m_0}=0.298^{+0.014+0.024}_{-0.013-0.022}$ and WMAP-9 data as $\Omega_{m_0}=0.295^{+0.016+0.026}_{-0.015-0.024}$ \cite{58}.
Other best fit values determined by  Strong gravitational lensing (SGL) and   CBS (CMB+BAO+SNIa) data   are $\Omega_{m_0}= 0.2891^{+0.0100}_{-0.0092}$,  $\Omega_{m_0}=0.2153^{+0.0078}_{-0.0058} $ \cite{59}, and the 5-year Supernova Legacy Survey (SNLS) data is  $\Omega_{m_0}=0.263\pm 0.042(stat)\pm0.032(sys)$ \cite{60}.  We first show preliminary results for which the matter
density  and the  Hubble parameter are  fixed at a constant values  of $\Omega_{m_0}=0.3$ \cite{40,61} and  $H_0 =72km/s/Mpc$ \cite{62}. The fact that
we have fixed $\Omega_{m_0}$ instead of marginalizing over it could
have artificially decreased somewhat the error bars of the
parameters.  \\
The first and simplest model to study is  matter dominated SCDM which is defined  in an anisotropic universe as follows:
\begin{equation}\label{18}
H^2=H_0^2\bigg(\Omega_{m_0}(1+z)^3+\Omega_{\sigma_0}(1+z)^6\bigg).
\end{equation}
We recall that the above equation is obtained from the first BI equation i.e. Eq. (\ref{10}).
The value of $\chi^2_{min}$ that minimize the $\chi^2(\Omega_{\sigma_0}, \theta)$ of Eq. (\ref{5}) (with $H(z)$ given by Eq. (\ref{18})) obtained by SNIa dataset. So we find
$\chi^2=\chi^2_{min}=468.994$ which results $\chi^2_{min}/dof=2.43$, where $dof$ is degrees of freedom\footnote{The value of $dof$ for the model equals the number of observational data points minus the number of parameters.}. Because the value of  $\chi^2_{min}/dof$ is  larger than 1. Therefore, SCDM is not suitable for SNIa data as shown in Fig. 1 (red line).\\
One of the simplest DE model is  the cosmological constant  $\Lambda$ with the  EoS parameter $w(z)=-1$. Although $\Lambda$ is the simplest model, it suffers from severe theoretical and conceptual problems, such as the fine-tuning and the cosmic coincidence problems \cite{48}.  In the  $\Lambda$CDM model Hubble's parameter is given by
 \begin{figure}[tb]
 \includegraphics[width=.45\textwidth]{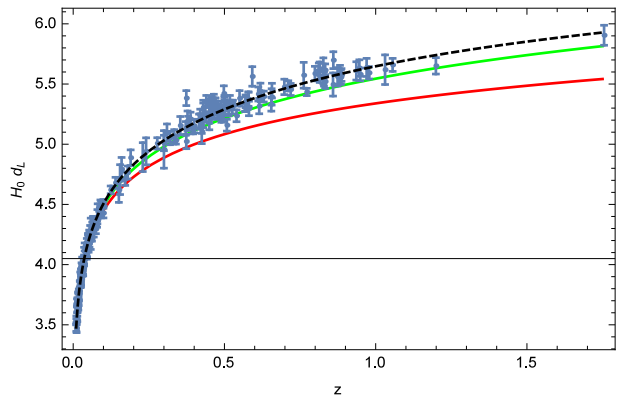}
\caption{The luminosity distance $  H_0 d_L$  versus the redshift $z$ for a flat BI  model.   The observational data points, shown with errorbars, are obtained from  SNIa \cite{47}. The  dashed line shows the  $\Lambda$CDM model and  the red solid line the SCDM model with  $\Omega_{\sigma_0}=0.013$, whereas     the green solid line shows the SCDM model with  $\Omega_{\sigma_0}=0$ (FRW model) \cite{61,64}.}
 \label{fig:1}
 \end{figure}
 \begin{equation}\label{19}
H^2=H_0^2\bigg(\Omega_{m_0}(1+z)^3+\Omega_{\sigma_0}(1+z)^6+(1-\Omega_{m_0}-\Omega_{\sigma_0})\bigg).
\end{equation}
 Using   Eq. (\ref{5}) we get the best fit value of  $\Lambda$CDM model parameter $(\Omega_{\sigma_0})$ is  $(0.0128)$ with $\chi^2_{min}=197.559$,
and the reduced $\chi^2$ value is  $\chi^2_{min}/dof=1.0236$. While  in the FRW model, the value of $\chi^2_{min}/dof$ is equal to $1.03$ \cite{40a,63}. This shows that the $\Lambda$CDM model with considering  the effects of the anisotropy, it taken to be the best-fit ones from SNIa data. In the case of  $1\sigma'$  errors on the value of
  $\Omega_{\sigma_0}=0.0128$ and by  solving  Eq. (\ref{7}) we get
  \begin{equation}\label{20}
\Omega_{\sigma_0}=0.0128\pm 0.00834.
\end{equation}
  In the following, we recall  for comparison   different DE models, we consider  $\Omega_{\sigma_0}=0.013$  with $\chi^2(\Omega_{\sigma_0}=0.013)=197.56$.
 Figure 1 illustrates the observational values of the luminosity distance $ d_L$ versus redshift $z$ together with the theoretical curves derived from (\ref{3}).
 The luminosity distance becomes larger when the cosmological constant is present and it  provides a good fit to
the data unlike to the case of SCDM model. In addition, Fig. 1 shows that a matter dominated universe (SCDM)  in a BI universe $(\Omega_{\sigma_0}=0.013)$  does not fit to the data.  For the case of SCDM model, a best-fit value of $\Omega_{\sigma_0}$  obtained in refs. \cite{ 40,61,64} is $\Omega_{\sigma_0}=0$ (isotropic universe). \\
 In order to fit the model with current observational data, we consider five DE models  in a flat BI in this section.
 \subsection{$w$CDM model with the constant EoS parameter  \cite{57}}
The next step is to allow for deviations from the simple $w=-1$ case, introducing a component with an arbitrary,
constant value for the EoS parameter. The accelerated expansion is achieved when $w<-1/3$ \cite{14}.  The Hubble parameter evolves according to the BI equation, which for  is
\begin{equation}\label{21}
H^2=H_0^2\bigg(\Omega_{m_0}(1+z)^3+\Omega_{\sigma_0}(1+z)^6+(1-\Omega_{m_0}-\Omega_{\sigma_0})(1+z)^{3(1+w)}\bigg).
\end{equation}
For $w=-1$  we recover the limiting form Eq. (\ref{19}).
Observational constraints on the $w$CDM model have been derived from many different data sets, hence it
provides a useful basis for comparing the discriminative power of different data. The currently preferred values of $w$ is given by: $w = -1.01 \pm 0.15$ \cite{65},  $w = -0.98 \pm 0.12$ \cite{6} and $w=-1.13^{+0.24}_{-0.25}$ from  the CMB and baryon acoustic oscillation (BAO) \cite{66}. Equation (\ref{21}) depends on two parameters, $\Omega_{\sigma_0}$ and $w$. Since the $\chi^2$  depends on two parameters. We use Eq. (\ref{5}) to produce and analyze the MLE.  In particular, for the case of $\Omega_{\sigma_0}=0.0128$ we obtain
 \begin{equation}\label{22}
\chi^2_{min}=\chi^2(-1.0146)=197.646,
\end{equation}
and the best fit value $w$ with the  $1\sigma'$ errors is
 \begin{equation}\label{23}
w=-1.0146\pm 0.0806.
\end{equation}
Also, for the value of $\Omega_{\sigma_0}=0.013$, we find $\chi^2_{min}=\chi^2(-1.0429)=197.276$ and  $\chi^2_{min}/dof=1.0221$. This indicates that  the minimization of  $w$CDM model,  exactly the   goodness-of-fit of the $\Lambda$CDM model for the same data.

 \subsection{Quiessence-$\Lambda$   ($q-\Lambda$) model \cite{40a} }
In this subsection we obtain a combination of cosmological constant with quiessence ($q-\Lambda$). Then we examine the effects of anisotropy on the cosmological implications of this model, and using the SNIa data we  probe observational constraints. The corresponding  form of $H(z)$ for the $q-\Lambda$ model in BI universe is
\begin{equation}\label{24}
H^2=H_0^2\bigg(\Omega_{m_0}(1+z)^3+\Omega_{\sigma_0}(1+z)^6+a_1(1+z)^{3(1+b_1)}+(1-a_1-\Omega_{m_0}-\Omega_{\sigma_0})\bigg).
\end{equation}
If we set $\Omega_{\sigma_0}=0.013$ and fit the 194 SNIa data, we get the best fitting values  as follows:
\begin{equation}\label{25}
\chi^2_{min}=\chi^2(a_1=0.00023,~b_1=3.1)=196.967.
\end{equation}
Considering the error bars we find that $a_1=2.3^{+1.3}_{-0.1}\times 10^{-4}$ and $b_1=3.1^{+0.421}_{-0.893}$. Based on Eqs. (\ref{16}) and (\ref{17}), the confidence levels of the best fit $w(z)$ and $q(z)$ calculated by using the    MLE for Eq. (\ref{5})  are plotted in Fig. 2 with  $\Omega_{\sigma_0}=0.013$ and $\Omega_{m_0}=0.3$.  From Fig. 2a, it is easy to see
that the   $w(z)$ can not  cross $-1$, which corresponds to the quintessence phase. On the other hand, $w(z)$ at $z\lesssim0.4$ behaves like the cosmological constant, i.e.  $w(z)=-1$. In conclusion, when our analysis is compared with that of ref. \cite{40a}, it is seen that the  EoS parameter
in the $q-\Lambda$ of FRW model has a lower slope than the EoS parameter in $q-\Lambda$ of  BI model, that this reflects that
anisotropy effects causes accelerating expansion so rapidly.  For a better insight, it was also seen that the EoS parameter  passes from  the dominant matter   to the DE region in the range of  $0.4\lesssim z\lesssim0.8$ \cite{40a}, while  it   happens in  the range of  $0.58\lesssim z\lesssim1.11$ in the present work  (see Fig. 1a).\\
 From Fig. 2b, we find out the
deceleration parameter is positive at large $z$, which indicates the earlier decelerating phase of the universe. Furthermore,  we can see that the best fit values of transition redshift and current deceleration parameter with confidence levels are  $z_t=0.404^{+0.021}_{-0.023}(1\sigma')^{+0.046}_{-0.042}(2\sigma')$ and $q_0(z)\sim-0.51$, which is consistent with the observations of \cite{67}. For comparison the combined analysis of SNe+CMB data with the $\Lambda$CDM model gives the range $z=0.69$ and the present value of the deceleration parameter $q_0=-0.55$ \cite{68} (assuming $\Omega_{m_0}=0.3$) and the matter dominated regime
($q(z)=1/2$) is reached by $z=1$. Recent studies have constructed  $q(z)$ takeing into account that the strongest evidence of accelerations happens at redshift of  $z\sim0.2$. In order to do so, the researcher  have  set $q(z)=1/2 (q_1 z+q_2)/(1+z)^2$ to reconstruct it and after that they have obtained  $q(z)\sim-0.31$ by fitting this model  to the  observational data \cite{69}.  Also it found that $q<0$ for $0\leqslant z \leqslant 0.2$ within the  $ 3\sigma'$ level.
Notice that the errors in Fig. 2 increase with redshift.\\
\begin{figure}
 \centering
 \subfloat[][]
{\includegraphics[scale=0.65]{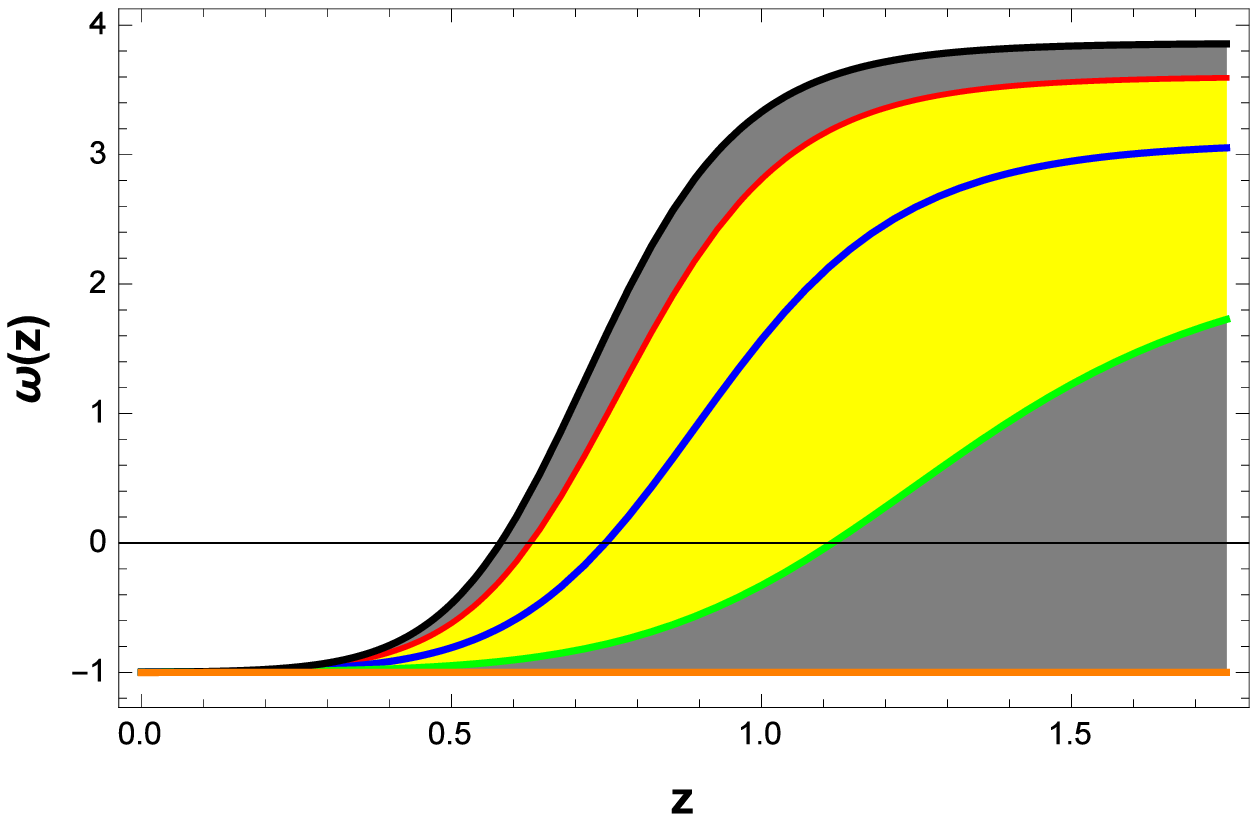}}~~~~
\subfloat[][]
 {\includegraphics[scale=0.65]{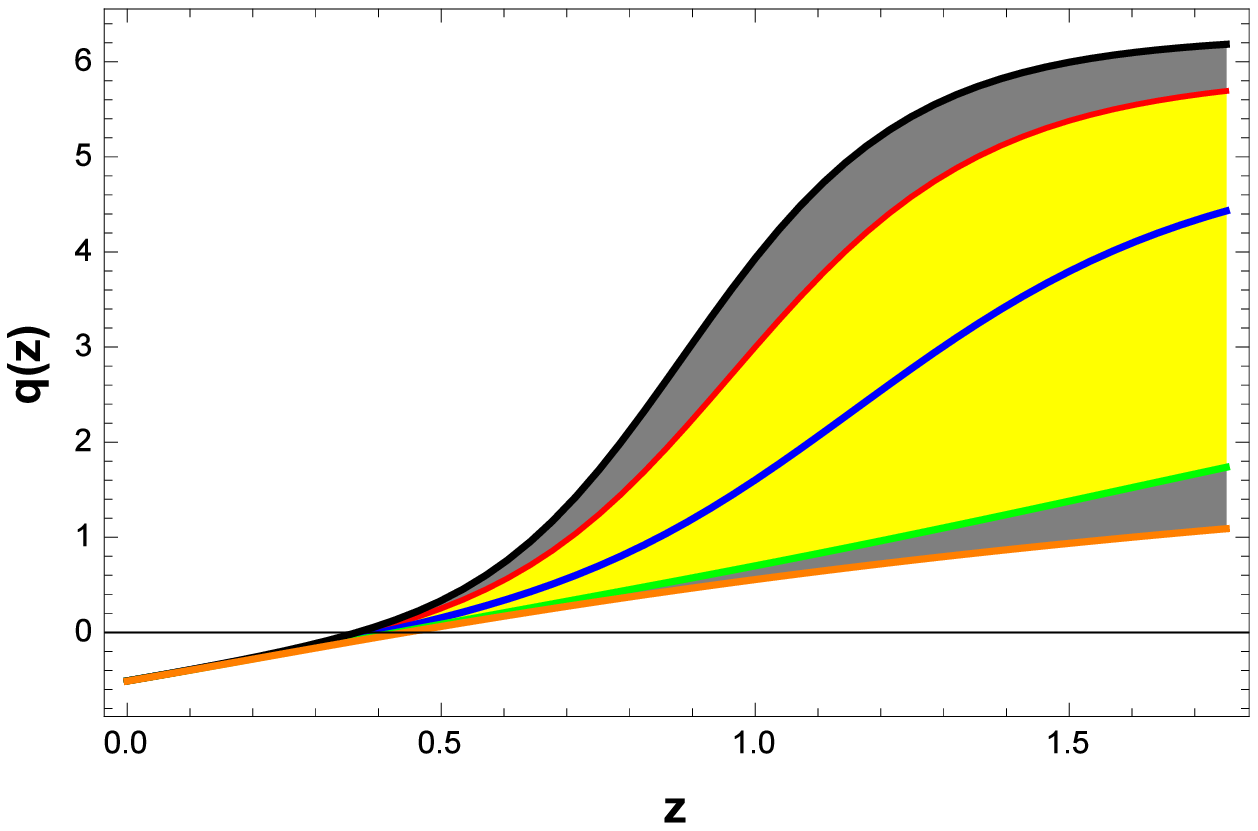}}~~~~
\caption{ The best fits of   $w(z)$ and $q(z)$  for the $q-\Lambda$ model in BI universe with $\Omega_{\sigma_0}=0.013$. The blue line is drawn
by using the best fit parameters.  The yellow and gray shaded areas show the  $1\sigma'$ and $2\sigma'$ errors   respectively. }
\label{fig:2}
\end{figure}
In the following, we follow the process of subsection A and   we introduce other DE models. Then we   obtain the form $H(z)$ and the parameters requiring fitting. Finally,
we figure out and minimize $\chi^{2}$ with respect to these parameters by considering $\Omega_{\sigma_0}=0.013$. In order to
simplify   and avoid confusion we only show the best fit curves without the corresponding $1\sigma'$ and $2\sigma'$ confidence limits.
 \subsection{ Linear parametrization    with  $w(z)=w_0+w_a z$ \cite{70,71,72,73,74} }
In order to discriminate between a cosmological constant and dynamical DE we use the DE EoS parameter parametrization \cite{70,71,72,73,74}
\begin{equation}\label{26}
w(z)=w_0+w_az,
\end{equation}
where $w_0$ and $w_a$ are the constant.  This model is interesting because it is described by simple field equations.
The cosmological constant ($\Lambda$CDM) corresponds to $w_0=-1$ and $w_a=0$, the case of constant EoS parameter ($w$CDM) corresponds to  $w_0=w= const$ and $w_a=0$, while the general case of time-evolving DE corresponds to $w_a\neq0$.  The DE density for this case is given by
\begin{equation}\label{27}
\rho_D(z)=\rho_0(z)e^{3w_a z} (1+z)^{3(1+w_0-w_a)}.
\end{equation}
In this case   Eq. (\ref{10}) gives the Hubble parameter
\begin{equation}\label{28}
H^2=H_0^2\bigg(\Omega_{m_0}(1+z)^3+\Omega_{\sigma_0}(1+z)^6+(1-\Omega_{m_0}-\Omega_{\sigma_0})(1+z)^{3(1+w_0-w_a)}e^{3w_a z}\bigg).
\end{equation}
We can then constrain the two parameters $w_0$ and $w_a$ by using SNIa data. The results obtained with standard rulers turned out to correspond well with previous work
by \cite{75}, whose results   were $w_0=-0.993 \pm 0.207$, $w_a=0.609 \pm 1.071$. As far as standard candles are concerned, the result of joint analysis from
WMAP+BAO+H0+SN given by \cite{76} is $w_0=-0.93 \pm 0.13$,  $w_a= 0.41^{+0.72}_{ -0.71}$.
 The allowed $1\sigma'$ confidence limits for ($w_0, w_a$), derived from the joint analysis SNIa data are: $w_0=-1.261^{+0.003}_{-0.024}$ and $w_a=1.417^{+0.147}_{-0.021}$ with $\chi^2_{min}=196.58$.
 \subsection{Chaplygin gas  and Generalized Chaplygin gas    }
 The expansion rate in the generalized Chaplygin gas (GCG) of BI model is governed by the equation
\begin{equation}\label{29}
H^2=H_0^2\bigg(\Omega_{m_0}(1+z)^3+\Omega_{\sigma_0}(1+z)^6+(1-\Omega_{m_0}-\Omega_{\sigma_0})[A+(1-A)(1+z)^{\alpha}]^{\frac{1}{2}}\bigg).
\end{equation}
It is straightforward to show that this equation of motion reduces to the isotropic universe in the limiting case corresponding to $\Omega_{\sigma_0}=0$ \cite{76a,77,78,79}.  Also, for the case of  $\alpha=6$ \cite{26,27,28,40a,80} the model recovers the Chaplygin gas model (CG).
 Within the framework of BI, we study a model based on   CG where our principal assumption is that the energy density $\rho_{ch}$ and pressure $p_{ch}$ are related by the following EoS
\begin{equation}\label{30}
 p_{ch}=-\frac{A}{\rho_{ch}}
\end{equation}
From Eq. (\ref{5}) it can be seen that the GCG behaves like pressureless dust at early times and like a cosmological constant during very late times.
In the following section, we will use the cosmic observations to constrain the GCG and CG models parameter $(A, \alpha)$. One can see that this constraint on parameter $\alpha$ is more stringent than the results in refs. \cite{81,82}, where the constraint results for the GCG model parameters are $A = 0.70^{+0.16}_{-0.17}$ and
$\alpha=-0.09^{+0.54}_{-0.33}$ at $2\sigma'$ confidence limits with the X-ray gas mass fractions of galaxy clusters and the dimensionless
coordinate distance of SNe Ia and FRIIb radio galaxies \cite{81}, and $A = 0.75\pm0.08$ and
$\alpha=-0.05^{+0.37}_{-0.26}$  at $2\sigma'$ confidence limits with the 115 SNLS SNe Ia data and the SDSS baryonic acoustic oscillations peak \cite{82}.
We allow now more anisotropy in our model and consider the
value for $\Omega_{\sigma_0}=0.013$. In the case of GCG model the best fit values for
$\alpha$ and $A$   are $[15.1^{+0.121}_{-0.092},0.999777^{+0.000223} _{-0.000448}]$ and the resulting $\chi^2_{min}$   is $196.791$ showing that there is an improvement in the quality of fit.   However, if we calculate the case of CG model for the
best fit among the  SNIa data (corresponding to a model with $A = 0.999\pm0.0225$) we find
$\chi^2_{min}=197.565$  (the errors are at the $1\sigma'$ level).
 \section{Comparison between the present models}
 \begin{figure}[tb]
 \includegraphics[width=.4\textwidth]{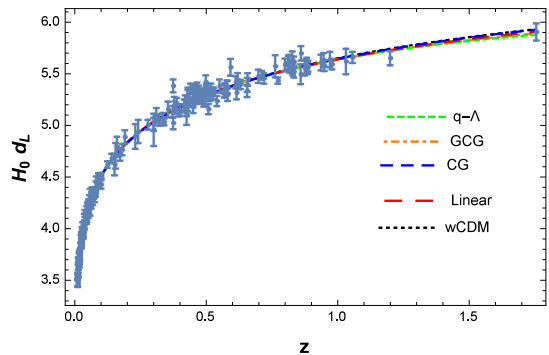}
\caption{Comparison  the luminosity distance $  H_0 d_L$ between various  DE models and the observational SNIa data \cite{47}  with  $\Omega_{\sigma_0}=0.013$ and $\Omega_{m_0}=0.3$.}
 \label{fig:3}
 \end{figure}
 \begin{widetext}
\begin{center}
\begin{table}
{ \bf Table I:} A comparison of the models used in the work. In all cases we have assumed   $\Omega_{m_0}=0.3$. \\
\begin{tabular}{cccccc}
\hline\hline  {\bf Model}        &{\bf $H(z)$}    &  {\bf p-test}& $\chi_{min}^2$ &$\chi_{min}^2/dof$& {\bf Best Fit Parameters} \\
\hline $\Lambda$CDM& $ H^2=H_0^2[ \Omega_{m_0}(1+z)^3+\Omega_{\sigma_0}(1+z)^6$
&&& \\  &$+(1- \Omega_{m_0}-\Omega_{\sigma_0})] $   &$-$ &$197.559$   &$1.0236$    &$\Omega_{\sigma_0}=0.0128\pm 0.00834$       \\
\hline SCDM & $H^2=H_0^2[\Omega_{m_0}(1+z)^3+\Omega_{\sigma_0}(1+z)^6]$ &$1$& $ 468.994$ &$2.4300$    &$-$       \\
\hline $w$CDM & $H^2=H_0^2[\Omega_{m_0}(1+z)^3+\Omega_{\sigma_0}(1+z)^6$
&&& \\  &$+(1-\Omega_{m_0}-\Omega_{\sigma_0})(1+z)^{3(1+w)}] $    &$0.405$&$197.276$   &  $1.0221$  &$w=-1.0146\pm 0.0806$\\
\hline q-$\Lambda$ &  $ H^2(z) = H_0^2 [\Omega_{m_0}(1+z)^3+\Omega_{\sigma_0}(1+z)^6$
&&& \\  & $+a_1(1+z)^{3(1+b_1)}+(1-a_1-\Omega_{m_0}-\Omega_{\sigma_0}) ]$    &$0.256$&$196.967$
&$1.0258$ &$a_1=2.3^{+1.3}_{-0.1}\times 10^{-4}$ ,
  $b_1=3.1^{+0.421}_{-0.893}$     \\
\hline Linear  & $H^2 (z)=H_0^2 [\Omega_{m_0} (1+z)^3+\Omega_{\sigma_0}(1+z)^6$ &&& \\ &
$+(1-\Omega_{m_0}-\Omega_{\sigma_0})(1+z)^{3(1+w_1-w_2)}e^{3 w_2 z}]$ &$0.387$&
$196.58$ & $1.0238$ &$w_0=-1.261^{+0.003}_{-0.024}$, $w_a=1.417^{+0.147}_{-0.021}$ \\
\hline GCG & $H^2 (z)=H_0^2 \{\Omega_{m_0}(1+z)^3+\Omega_{\sigma_0}(1+z)^6$
&&& \\  & $+[1-\Omega_{m_0}-\Omega_{\sigma_0}][A+(1-A)(1+z)^{\alpha}]^{\frac{1}{2}}\}$ &$0.319$&$196.791$ & $1.0401$ & $\alpha=15.1^{+0.121}_{-0.092}, A=0.999777^{+0.000223} _{-0.000448}$ \\
\hline CG & $H^2 (z)=H_0^2 \{\Omega_{m_0}(1+z)^3+\Omega_{\sigma_0}(1+z)^6$
&&& \\  & $+[1-\Omega_{m_0}-\Omega_{\sigma_0}][A+(1-A)(1+z)^{6}]^{\frac{1}{2}}\}$ &$0.062$&$ 197.565$ & $1.0237$ & $A = 0.999\pm0.0225$      \\
\hline\hline
\end{tabular}
\end{table}
\vspace{-0.2cm}
\end{center}
\end{widetext}
 In this section, we choose several popular DE models and estimate their best fitted
parameters using the SNIa data. We also examine consistency of our findings with
other independent results from the literature and  we apply the effects of anisotropy on  the p-test to rank 6 representative parametrizations of $H(z)$. The comparison of the parametrizations considered is shown in Table I.   In all cases we have assumed priors corresponding to flatness and $\Omega_{\sigma_0}=0.013$ and $\Omega_{m_0}=0.3$. It can be seen that the linear parametrization  model has the smallest  $\chi_{min}^2$. Also the $\chi_{min}^2$ per degree of freedom for
the best-fit for the different cases is given in Table I.
 Figure  3  shows the  luminosity distance measured by the SNIa compared to various possible models of Table I, which strongly favours a DE model.
 It  shows that  the effect of anisotropy on $H_0 d_L$ is nearly identical for the set of models as individual curves can hardly be distinguished.
 In the next step, we examine the EoS  and deceleration parameter for the DE models studied in the previous section. In Fig. 4 we plot the best fit $w(z)$ and $q(z)$ for 5 representative parametrizations of Table I.  Among the models, the linear parametrization and $w$CDM models that are providing the best
fit to the EoS parameter $w(z)$ exhibit crossings of the  $w(z)=-1$ divide line (see Fig. 4a). Fig. 4b shows that the universe transits from a matter dominated epoch at early times to the acceleration phase in the present time, as expected. In Table II, we list the results of $w$, $q$ and  $z_t$  for the current universe ($z=0$).
We have also found that the constraints obtained on the parameter
values by the SNIa datasets. The Gaussian distribution is the most important distribution in statistics, because it is so ubiquitous
(so ``normal''), appearing in many different experimental settings, and because many other distributions approach the normal distribution as soon as they become ``messy''. The full probability distribution, for example, converges to the normal distribution for a large numbers of data samples.
The central limit theorem of probability theory tells us that a sum of identically distributed independent variables has, in the limit, a normal distribution.
It is seen from the likelihood plots (Fig. 5) that the likelihood functions are well fitted to a Gaussian distribution function for each
dataset.  The corresponding constraints on model parameters are summarized in Table I.\\
Finally, the reduced form of  $H(z)$ compared to $\Lambda$CDM model determined as
\begin{equation}\label{30}
 H'^2(z)=\frac{H^2(z)-H^2_{\Lambda CDM}(z)}{H^2_0},
\end{equation}
where $H'^2(z)$ is the reduced form of  $H(z)$ and $H^2_{\Lambda CDM}(z)=0.7+0.3(1+z)^3$. In the Fig.  6, we plot the deviation of the squared
Hubble parameter $H^2/H^2_0$ from $\Lambda$CDM over redshift for the best fit.   From this figure, we see that  expansion is faster when DE  is $q-\Lambda$ model as compared to other DE models. But   in range of $z<0.5$, the curves are coincide,  that is  the effect of anisotropy parameter density is constant.

\begin{figure}
 \centering
 \subfloat[][]
{\includegraphics[scale=0.35]{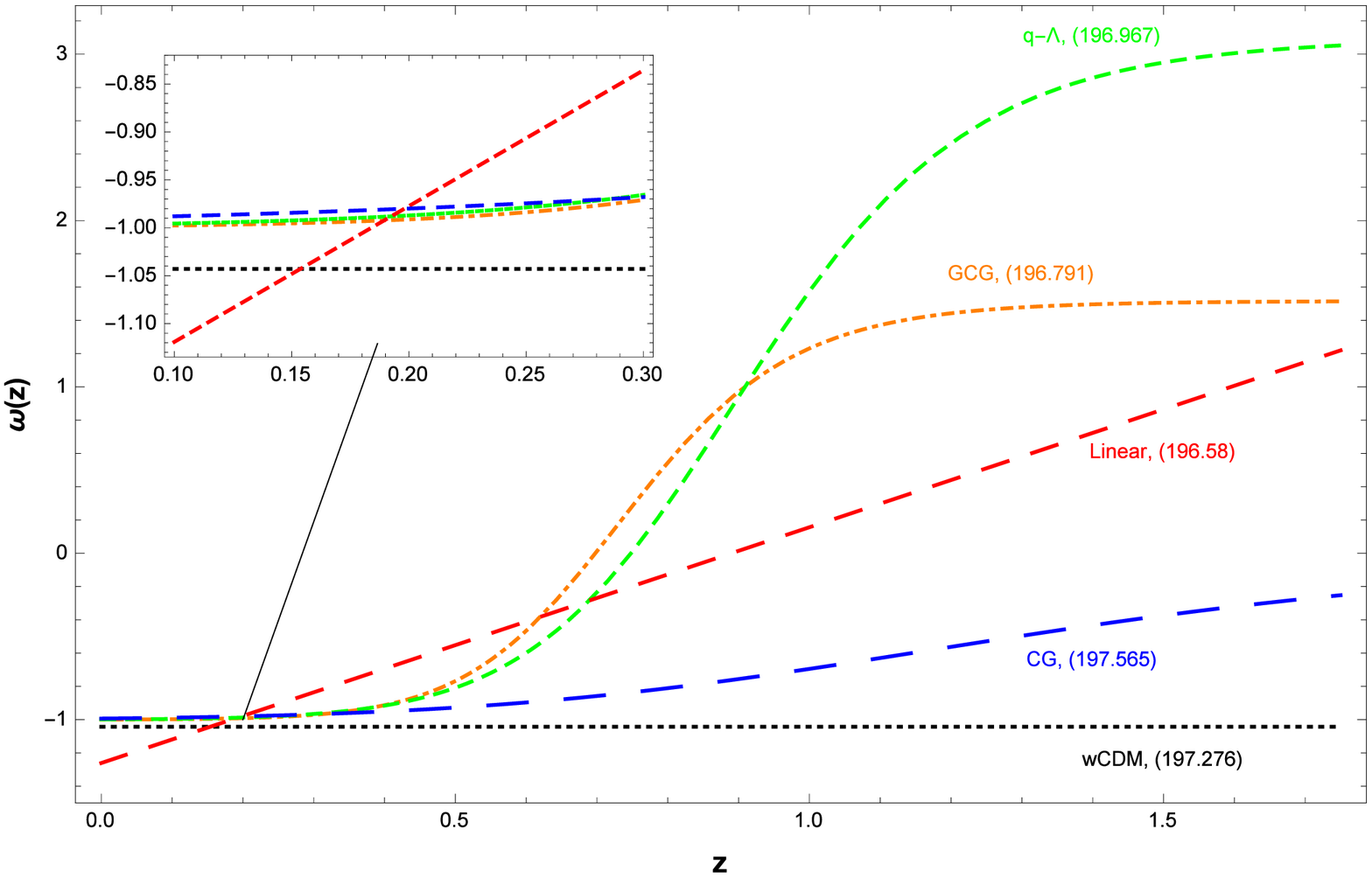}}~~~~
\subfloat[][]
 {\includegraphics[scale=0.35]{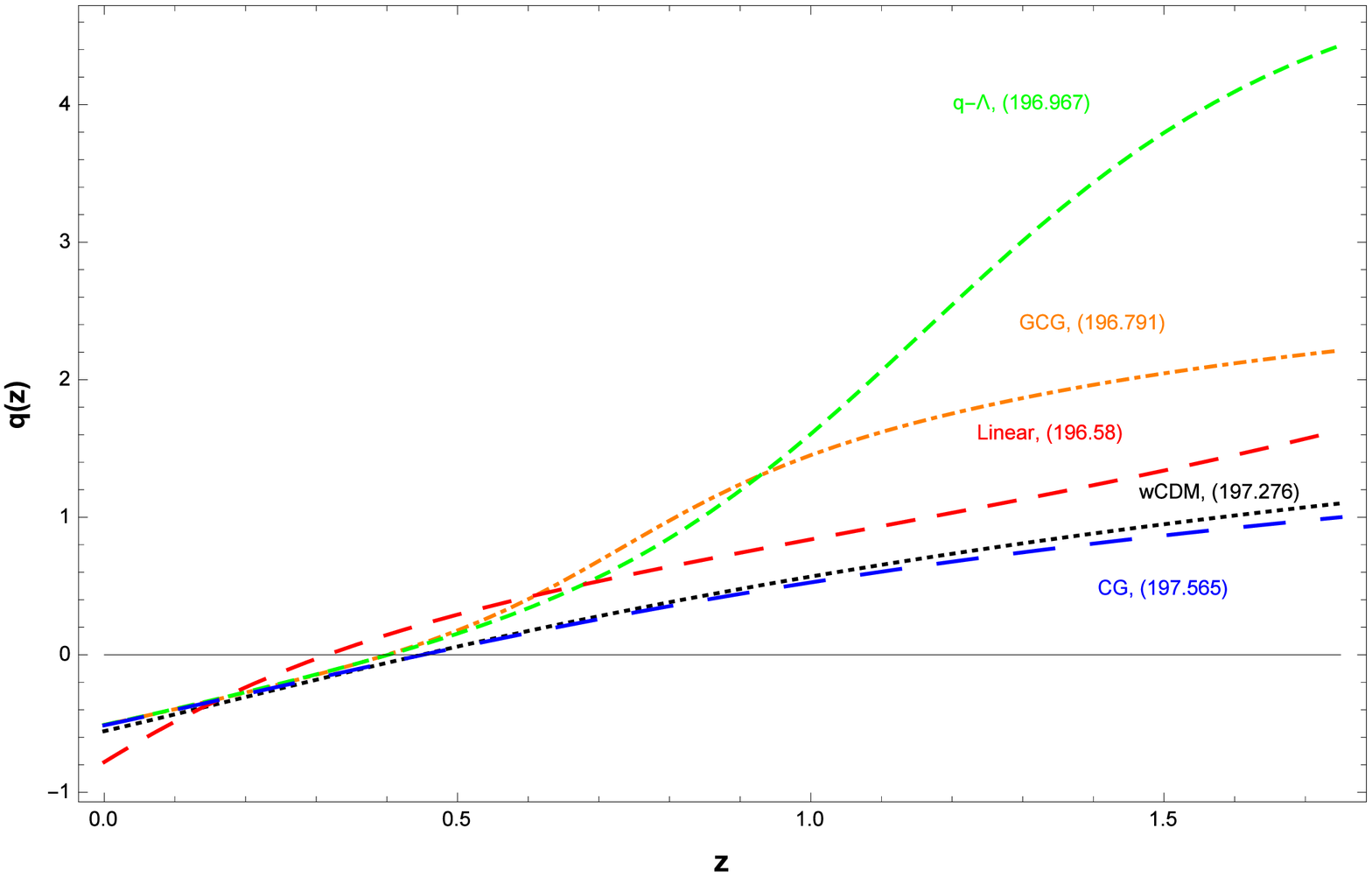}}~~~~
\caption{The plots of $w(z)$ and $q(z)$  for some representative parametrizations  in BI universe with $\Omega_{\sigma_0}=0.013$  for the cosmological models of Table I with the SNIa  datasets.  Notice that the phantom divide line  crossing at best fit occurs only for  the linear parametrization and $w$CDM models.  The numbers in the  parentheses represent the value of $\chi_{min}^2$ for each DE models. Notice that they all cross the line $w(z)=-1$ also known as the phantom divide line. }
\label{fig:4}
\end{figure}
\begin{widetext}
\begin{center}
\begin{table}
{ \bf Table II:} The results of $w(z=0)$, $q(z=0)$  and  $z_t$ for five DE models    used in the work. \\
\begin{tabular}{cccccc}
\hline\hline  {\bf Model}~&{\bf$w$CDM  } ~  &  {\bf Linear} ~& {\bf ~ q-$\Lambda$}~ ~&{\bf GCG} ~~& {\bf CG} \\
\hline $w(z=0)$&$-1.043$ &$-1.261$   &$-0.998$    &$-0.999$ &$-0.993$      \\
\hline $q(z=0)$ & $-0.555$& $ -0.781$ &$-0.509$    &$-0.511$  &$-0.513$        \\
\hline $z_t$ & $~~0.449$&$~~0.315$   &  $~~0.402$  &$~~0.400$&$~~0.451$\\
\hline\hline
\end{tabular}
\end{table}
\vspace{-0.2cm}
\end{center}
\end{widetext}
\begin{figure}
 \centering
 \subfloat[][]
  {\includegraphics[scale=0.3]{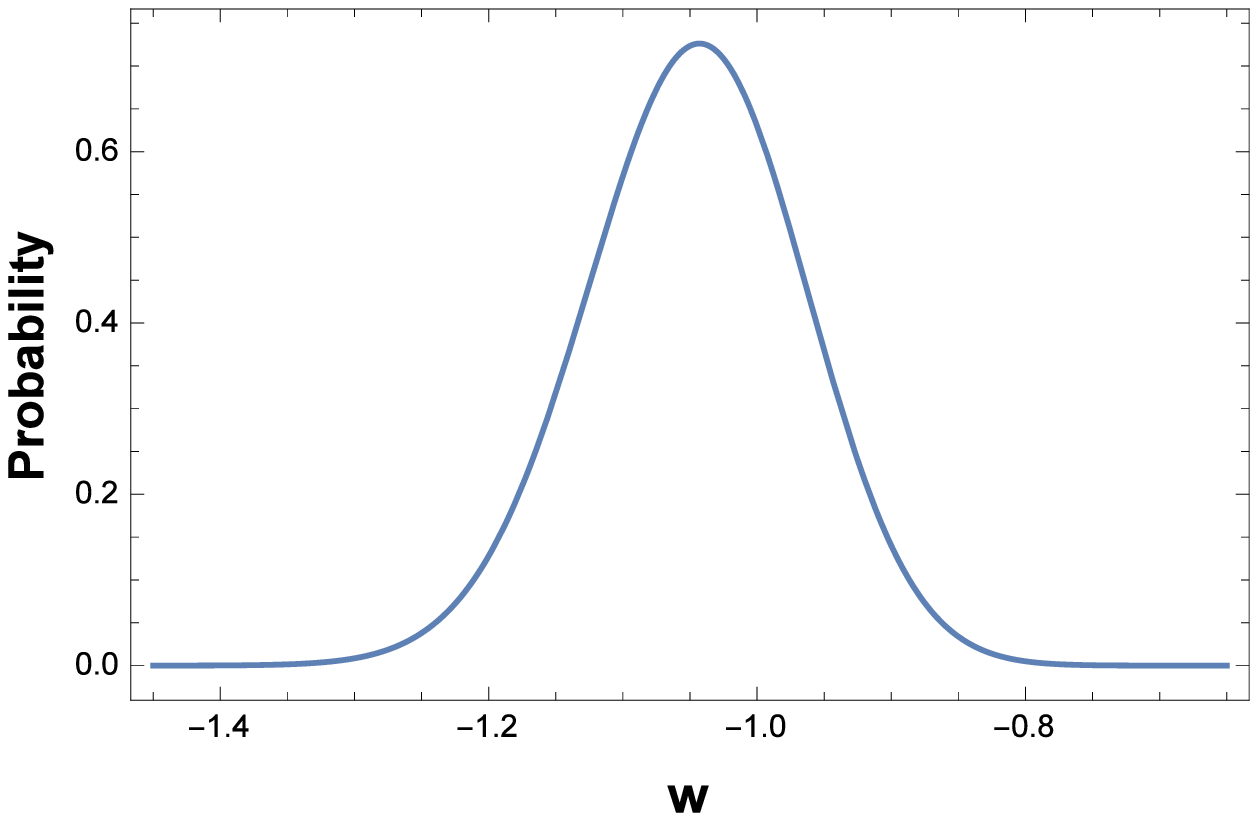}}~
   \subfloat[][]
{\includegraphics[scale=0.3]{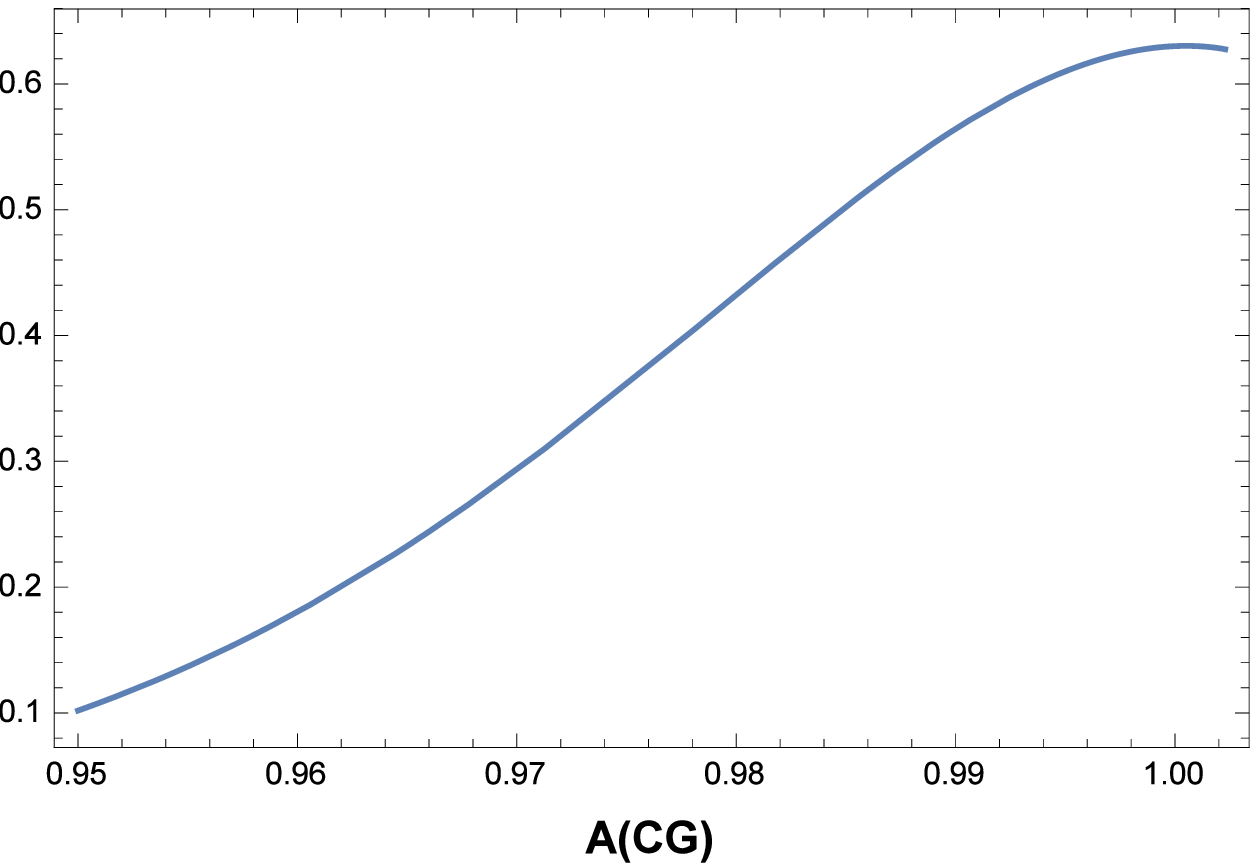}}~
\subfloat[][]
 {\includegraphics[scale=0.3]{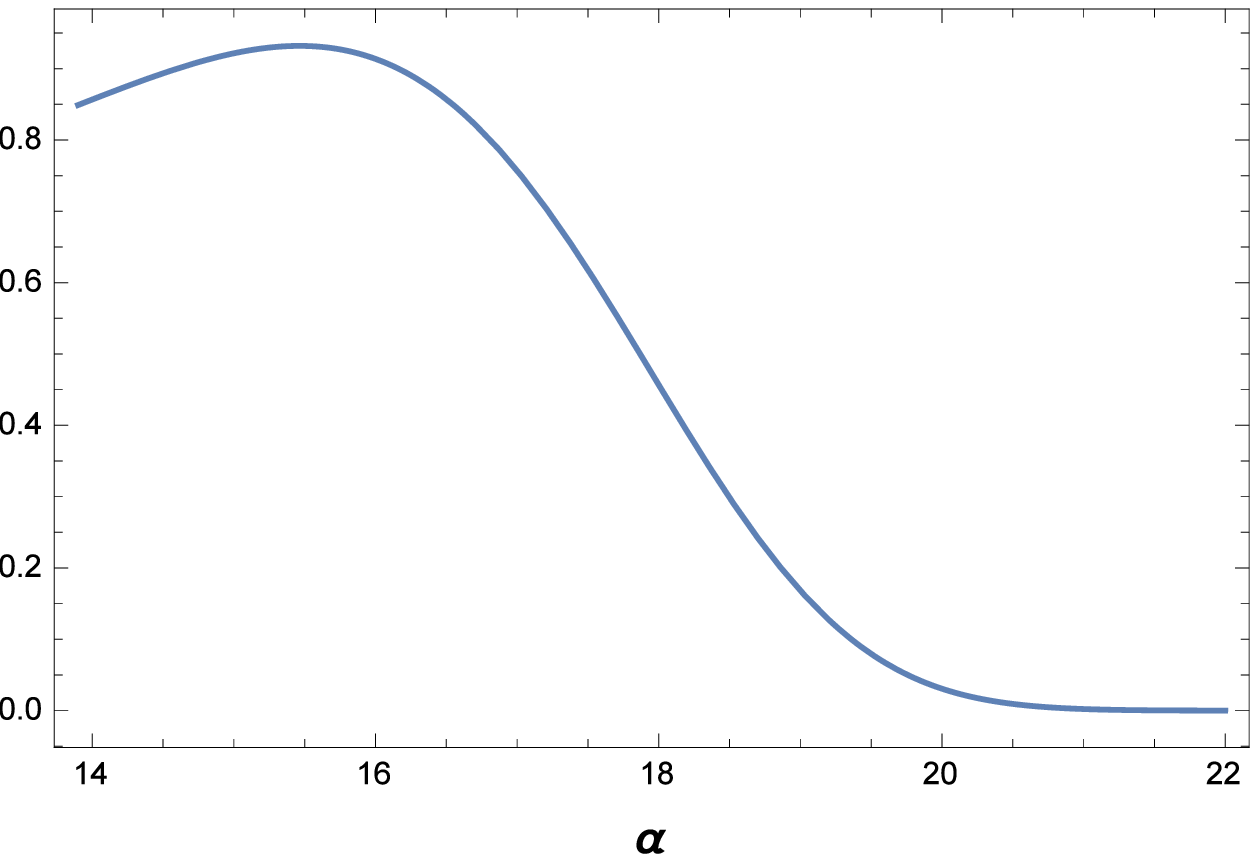}}~
 \subfloat[][]
 {\includegraphics[scale=0.3]{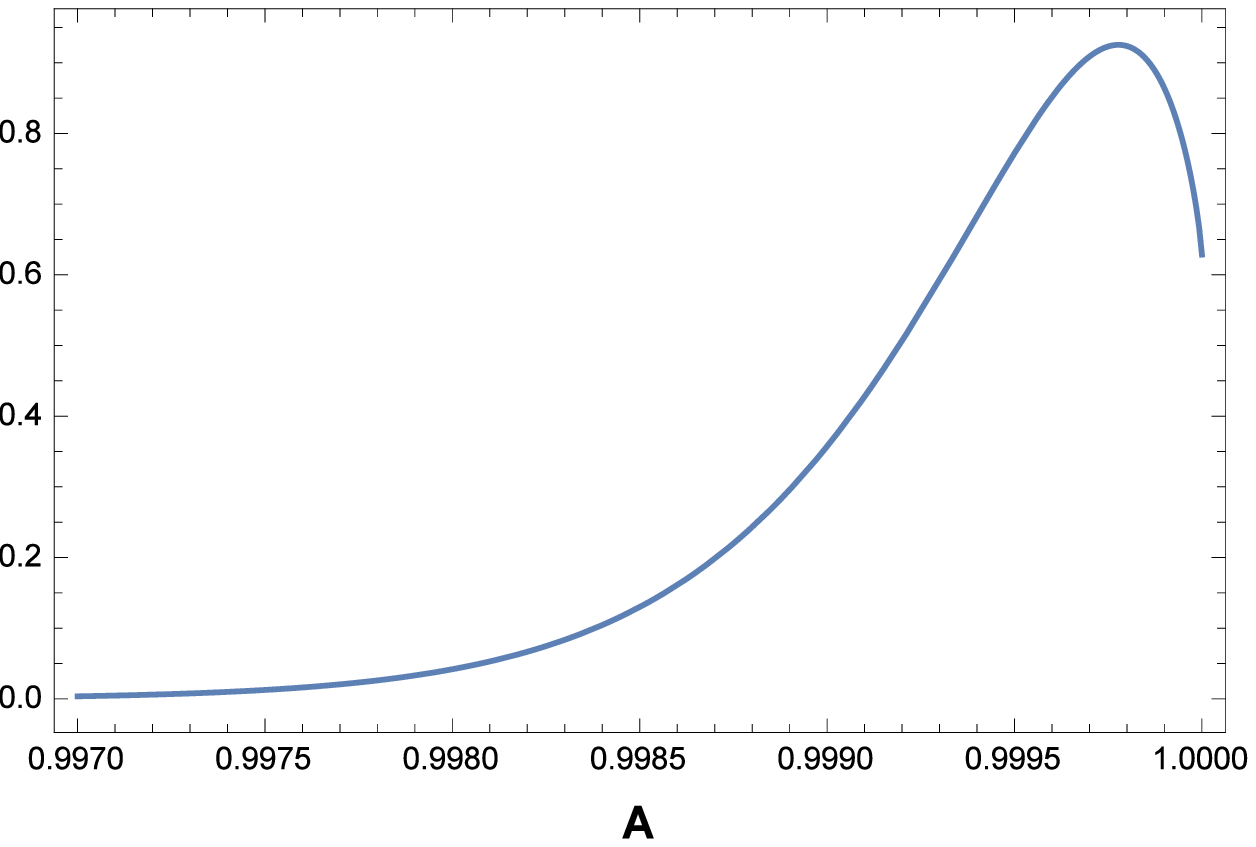}}\\
 \subfloat[][]
 {\includegraphics[scale=0.3]{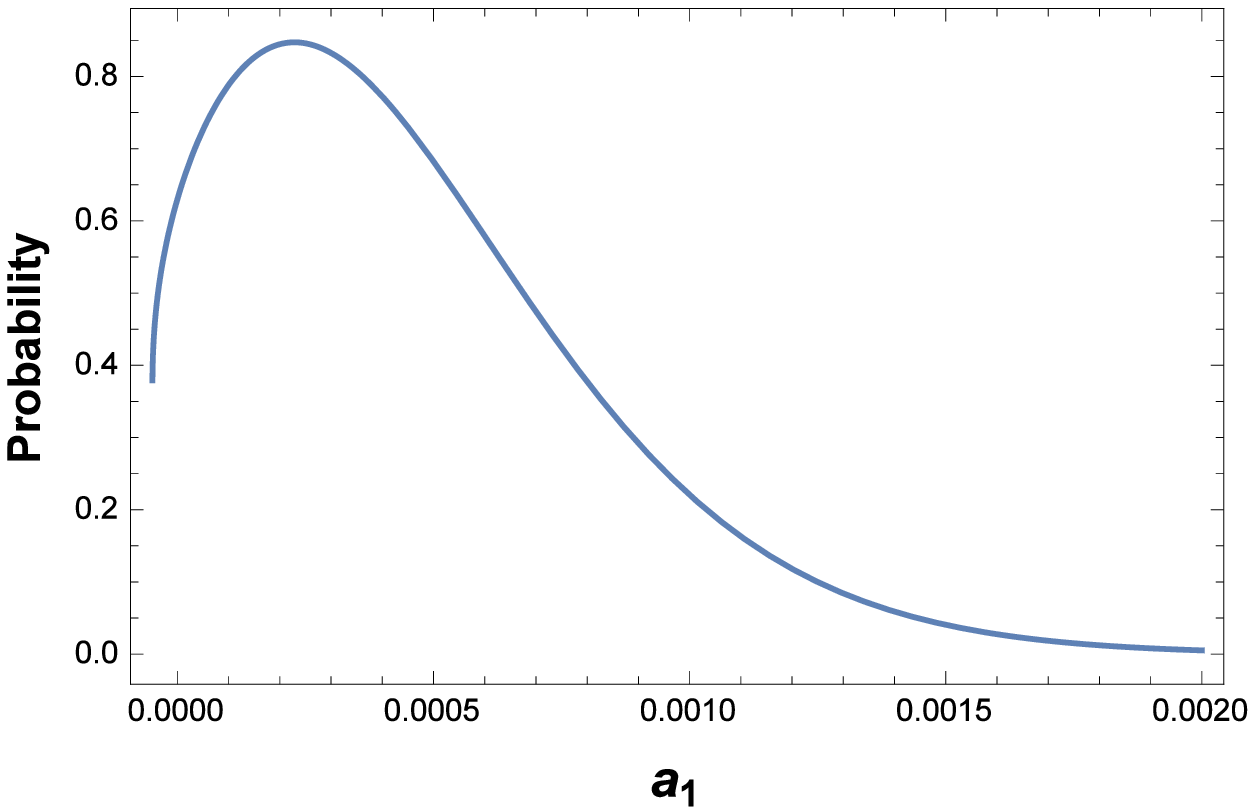}}~
 \subfloat[][]
 {\includegraphics[scale=0.3]{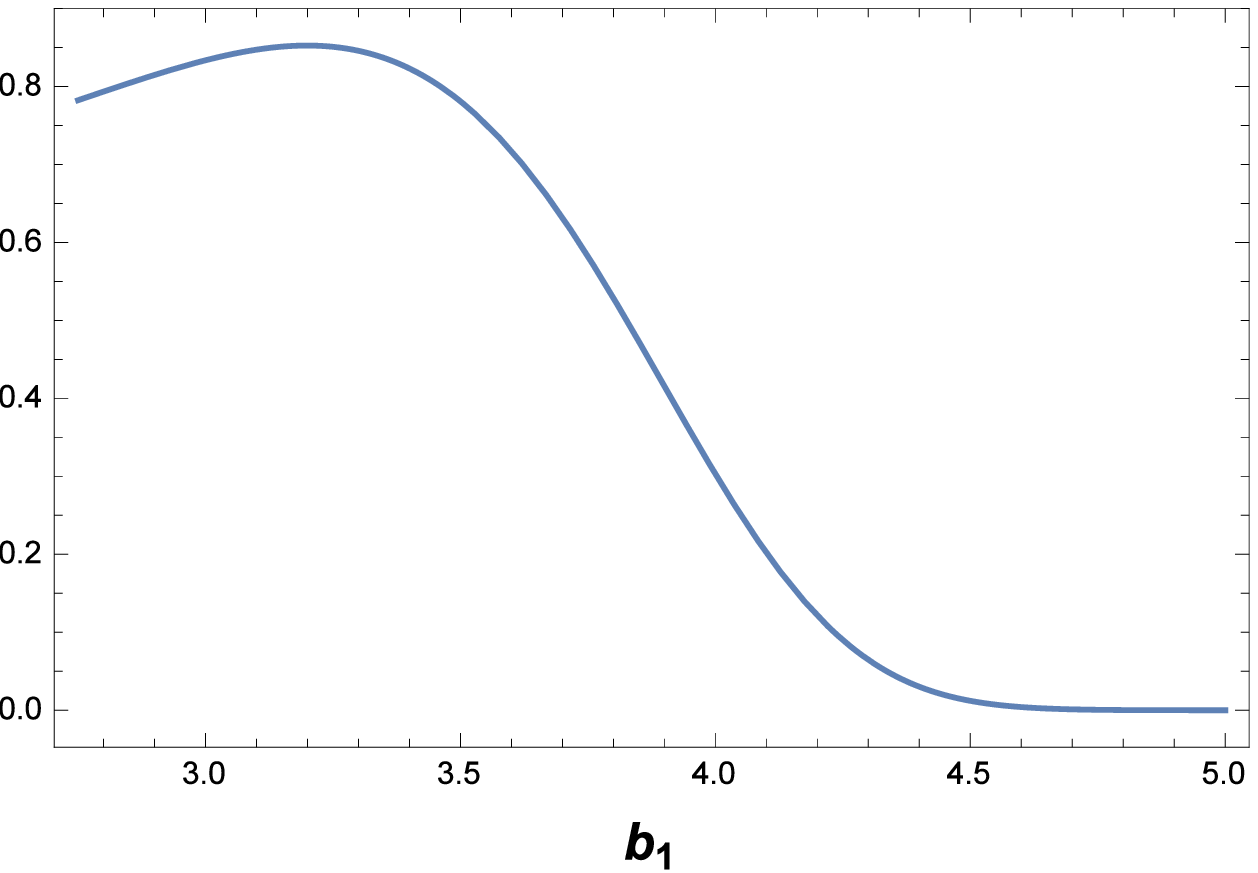}}~
  \subfloat[][]
 {\includegraphics[scale=0.3]{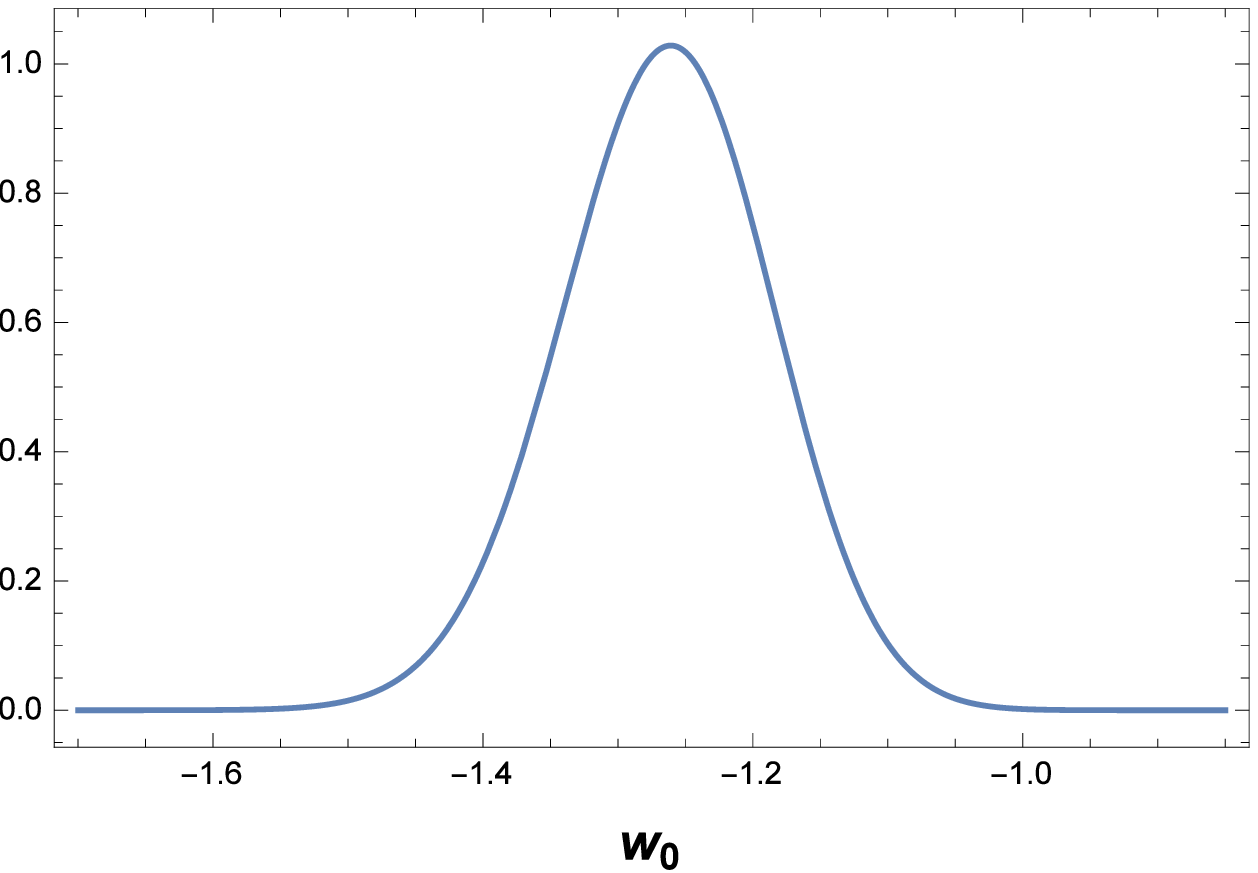}}~
 \subfloat[][]
 {\includegraphics[scale=0.3]{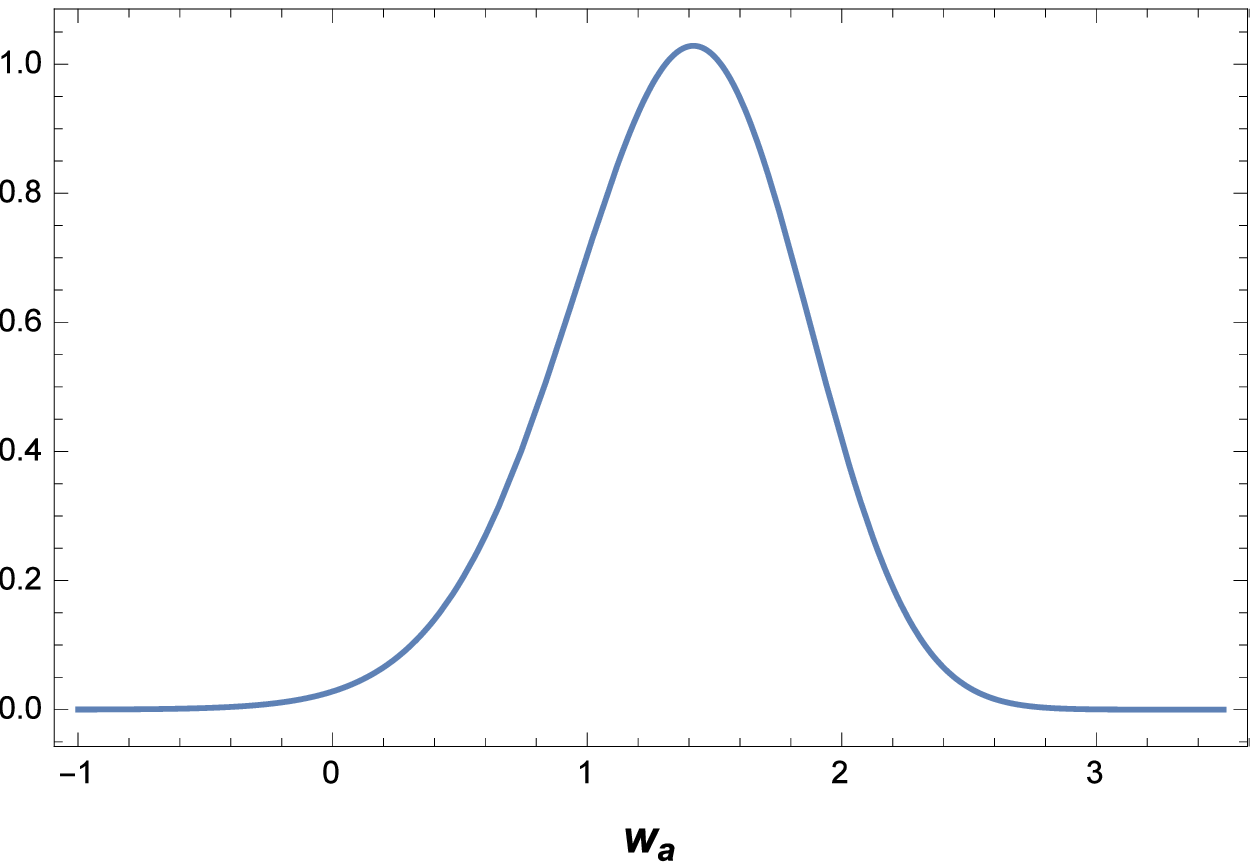}}
\caption{ The marginalized probabilities of the various models  are shown in Table I. The  curves are the result from SNIa data.  }
\label{fig:5}
\end{figure}
\begin{figure}[tb]
 \includegraphics[width=.4\textwidth]{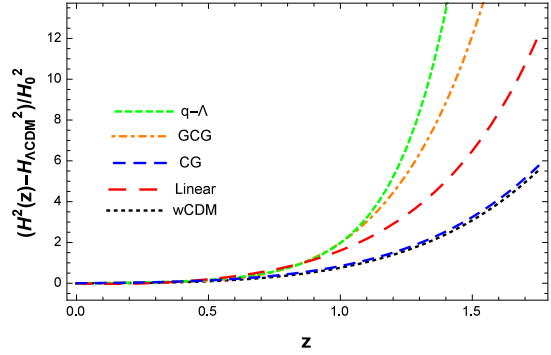}
\caption{The reduced Hubble parameter $H'^2(z)$ for some of the best fits of the cosmological models of Table I  with $\Omega_{\sigma_0}=0.013$.  }
 \label{fig:6}
 \end{figure}
\section{Light curve parameters of SNLS $SNe~ Ia$ }
 To derive the brightness, light-curve shape and SN color estimates required for the cosmological analysis, the time sequence of photometric measurements for each SN was fit using a SN light-curve model.   Guy \textit{et al}. \cite{83} presents the light curves of the SNLS
SNe Ia themselves, together with a comparison of SN light curve fitting techniques, color laws, systematics and parameterizations. Conley \textit{et al}. \cite{83a} compared the performances of different light-curve fitters while also introducing their own empirical fitter, SiFTO, and concluded that SALT2 along with SiFTO
perform better than both SALT (which is conceptually different from its successor SALT2) and MLCS2k2 (Jha \textit{et al}. \cite{83b}) when
judged by the scatter around the best-fit luminosity-distance relationship.   We parameterize the SN Ia light curves for distance estimation using the combined results from updated version of  light curve fitter   SiFTO. It technique provide an estimate of the SN peak rest-frame B-band apparent magnitude at the epoch of maximum light in that filter, a measure of the SN light curve
shape, and an estimate of the SN optical B-V color ($\mathcal{C}$).  Guy \textit{et al}. \cite{83} discarded from the SNLS catalog 252 SNe with a peak rest-frame $(B-V)>0.2$.  This cut, applied to both SALT2 and SiFTO samples, discards 11 SNe. There are also three SNe whose peak magnitudes could not be obtained with SiFTO due to a lack of observations in the $g_M$ and $z_M$ bands. Hence, that leaves $N=238$ SNLS SNe.    From the fits to the light-curves, we computed a
rest-frame-B magnitude, which, for perfect standard candles, should vary with redshift according to the luminosity distance.
This rest-frame-B magnitude refers to observed brightness, and therefore does not account for brighter-slower and brighterbluer correlations \cite{84}. The observed distance modulus of each SN is given by \cite{83,84,85}
 \begin{equation}\label{31}
 \mu_B=m_B+\alpha(s-1)-\beta \mathcal{C}-M_B ,
\end{equation}
where $m_B$  is the observed peak magnitude in rest frame B band, $s$ is the stretch (a measure of light-curve shape), $\mathcal{C}$ corresponds to the supernova color at maximum brightness. Notice that $\alpha$, $\beta$ and $M_B$ are nuisance parameters in the distance estimate, which should be fitted simultaneously with
the cosmological parameters. The theoretical model parameters are determined by
minimizing the quantity \cite{83,84,85}
\begin{equation}\label{32}
\chi^2=\sum^N_{i=1}\frac{(\mu_{B,i}-\mu_{th}(z))^2}{\sigma'^2(\mu_B,i)+\sigma'^2_{int}},
\end{equation}
where  $\mu_{th}$ is given by Eq. (\ref{2}),  $N=238$ for the  joint light-curve analysis (JLA) SNIa,  $\sigma'^2(\mu_B,i)$ and $\sigma'^2_{int}$ are the errors due to flux uncertainties, intrinsic dispersion of SNIa absolute magnitude. The SN-specific dispersion $\sigma'^2(\mu_B,i)$ is defined by
\begin{equation}\label{33}
\sigma'^2(\mu_B,i)=\sigma'^2_{m_B,i}+\alpha^2\sigma'^2_{s,i}+\beta^2\sigma'^2_{\mathcal{C},i}+C_{m_Bs\mathcal{C},i},
\end{equation}
Where $\sigma'^2_{m_B,i}$, $\sigma'^2_{s,i}$ and $\sigma'^2_{\mathcal{C},i}$ are the standard errors of the peak magnitude and light-curve parameters of the i-th SN.
The term $C_{m_Bs\mathcal{C},i}$ comes from the covariances among $m_B, ~s,~ \mathcal{C}$  and
likewise depends quadratically on  $\alpha$ and $\beta$. For the case of $\sigma'^2_{int}$, we perform a first fit with an initial value
(typically 0.15 mag.), and then calculate the $\sigma'^2_{int}$ required to
obtain a reduced $\chi^2=1$. We then refit with this more accurate
value. We fit 6 cosmologies to the data that was introduced in Table I. To find the best-fit coefficients $\alpha$, $\beta$ and $M_B$ and the
cosmological parameters that define the fitted model, we use MLE techniques, which is based on the maximization of a joint
likelihood function. Using the MLE approach, we find
that the best-fit curvature parameter using the JLA SNe Ia data  and the  considering effect of anisotropy, the optimized nuisance parameters are  $\Omega_{\sigma0}=-0.003\pm0.033$, $\Omega_{m0}=0.28\pm 0.202$, $\alpha=1.4022\pm0.111$, $\beta=3.279\pm0.145$ and $M_B=-19.1\pm0.067$ with $\sigma'_{int}=0.104$.  In Figure 7 we show the (normalized) likelihood
distribution for each parameter ($\Omega_{\sigma0}$, $\Omega_{m0}$, $\alpha$, $\beta$, $M_B$), according to
the factor $e^{-\chi^2/2}$ , and $1\sigma'$, $2\sigma'$ contours for the joint
distribution of each pair of parameters. For each parameter, the
likelihood distribution is well approximated by a Gaussian, and
the stated confidence interval is a 68\% (i.e., $\pm1\sigma'$) interval for this Gaussian.  The
maximum value of the joint likelihood function for an anisotropic universe  corresponds to $-2 \ln L=-230.656$.  In conclusion, when our analysis is compared with that of ref. \cite{85}, it is seen that the maximum value of the joint likelihood function in the $\Lambda$CDM of
FRW model has a bigger than  the maximum value of the joint likelihood function  in the $\Lambda$CDM of BI model, that imply that anisotropy effects
causes  improves the data fitting  in the $\Lambda$CDM DE model.  The resulting color and light-curve shape corrected peak B-band magnitudes for    the various cosmological models are presented in   Table III. The parameters $\alpha$, $\beta$, and $M_B$ are nuisance parameters which are
fitted simultaneously with the cosmological parameters.  From   Table III,  it can be observed that the $\Lambda$CDM DE model has the smallest
 $\chi_{min}^2/dof$.
 \begin{figure}[tb]
 \includegraphics[width=.13\textwidth]
 {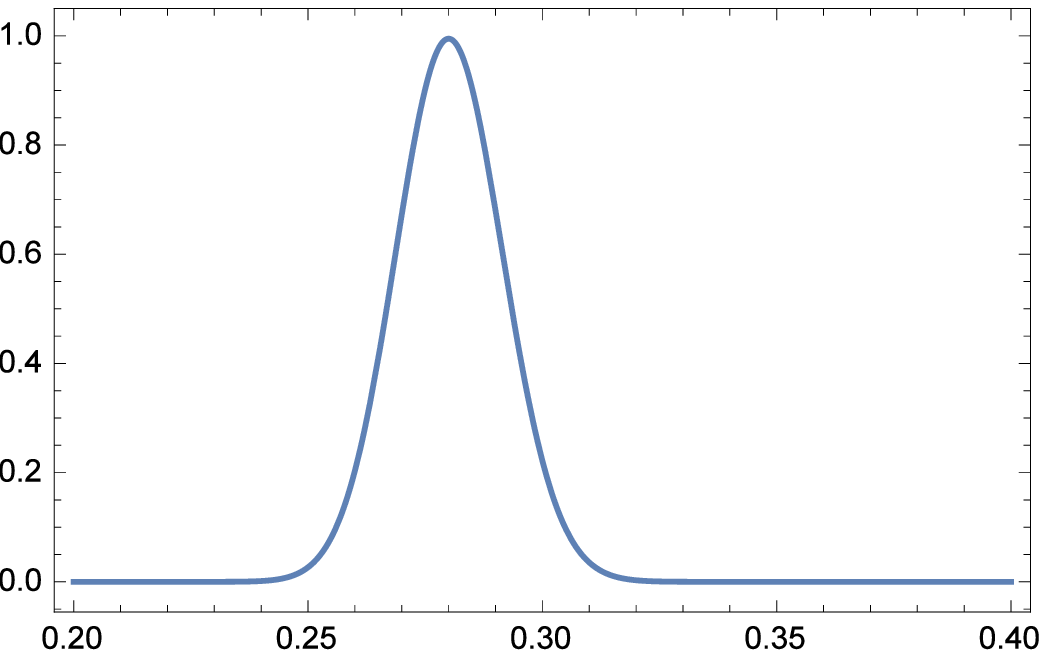}\hspace{.00001cm}~~~~~~~~~~~~~~~~~~~~~~~~~~~~~~~~~~~~~~~~~~~~~~~~~~~~~~~~~~~~~~~~~~~~~~~~~
 ~~~~~~~~~~~~~~~~\\
 \includegraphics[width=.14\textwidth]{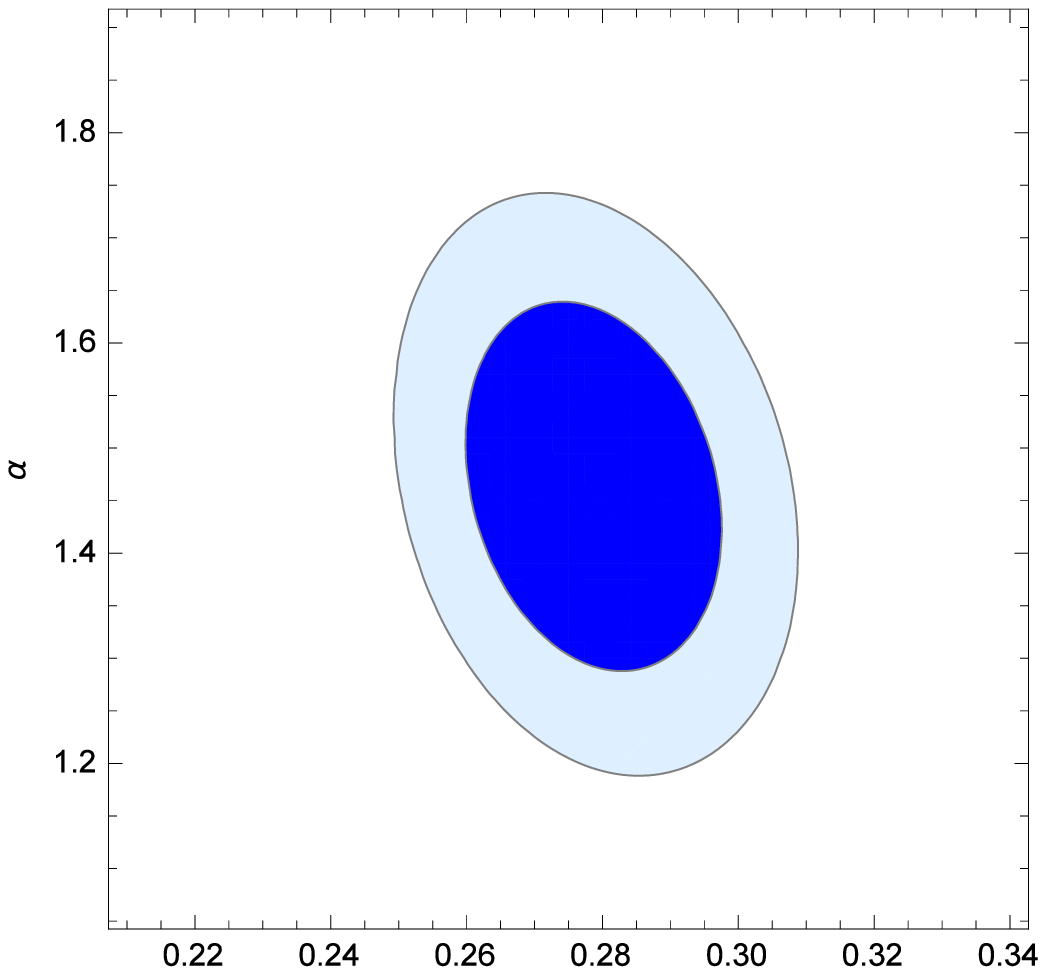}\hspace{.00001cm}
 \includegraphics[width=.13\textwidth]{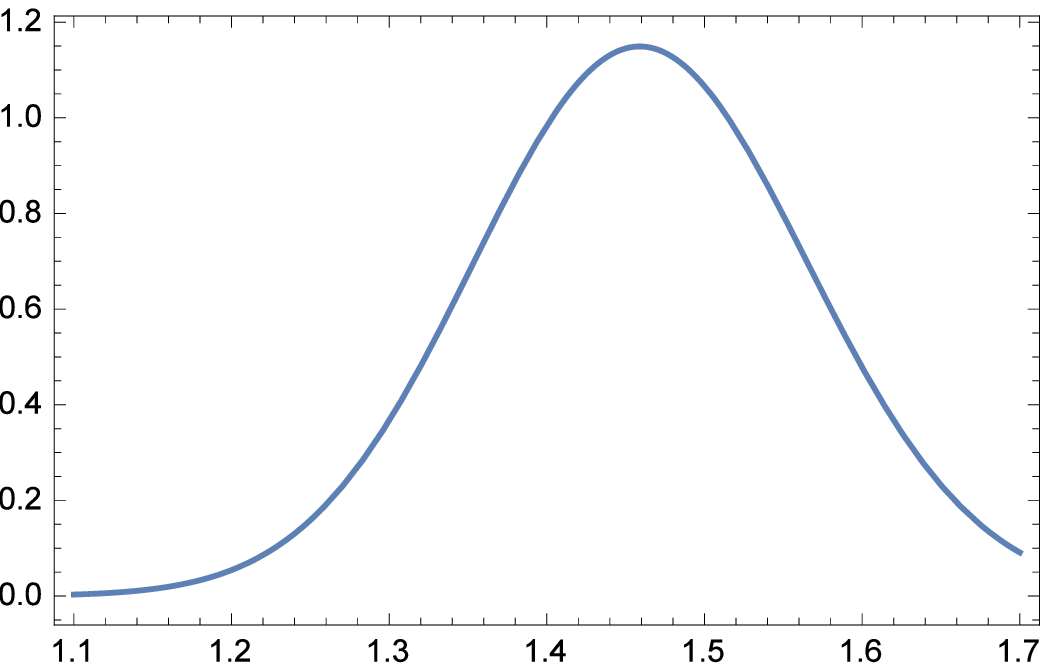}\hspace{.00001cm}~~~~~~~~~~~~~~~~~~~~~~~~~~~~~~~~~~~~~~~~~~~~~~~~~~~~~~~~~~~~~~~~~~~~\\
 \includegraphics[width=.14\textwidth]{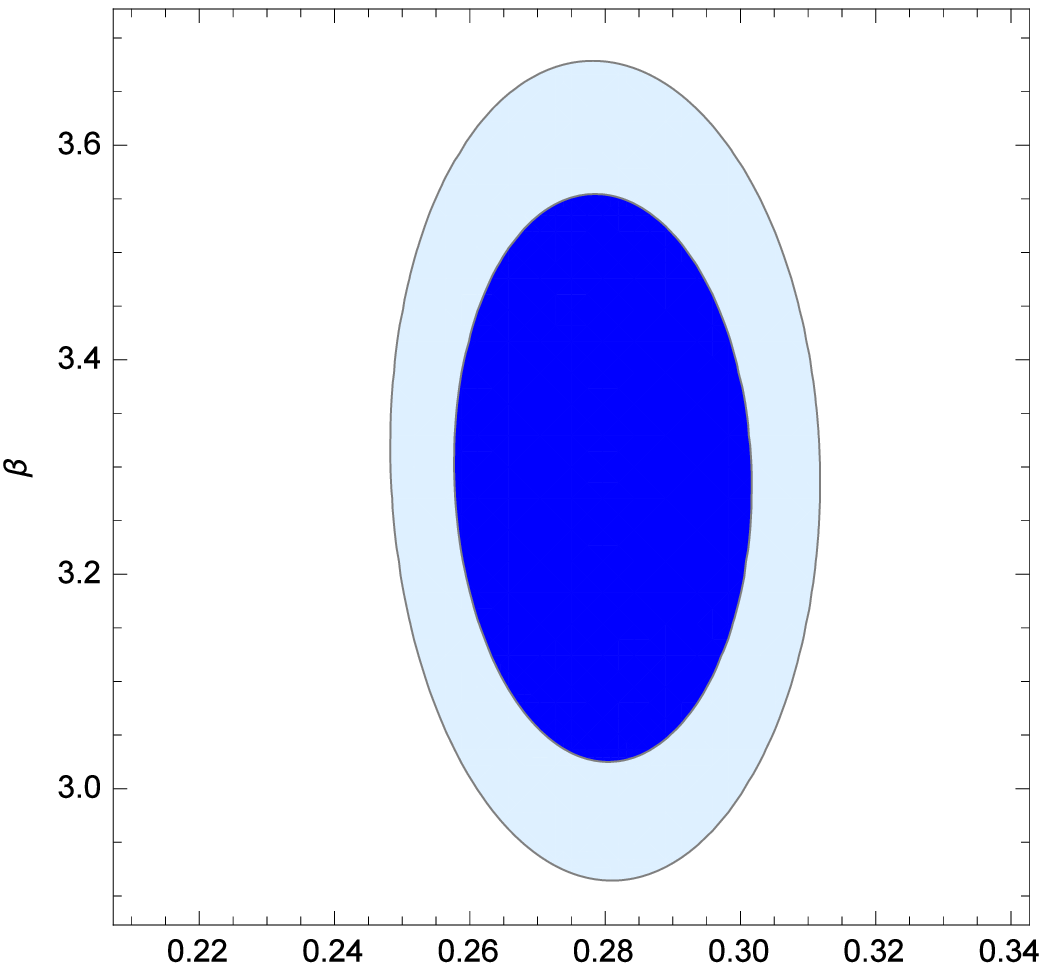}\hspace{.00001cm}
\includegraphics[width=.13\textwidth]{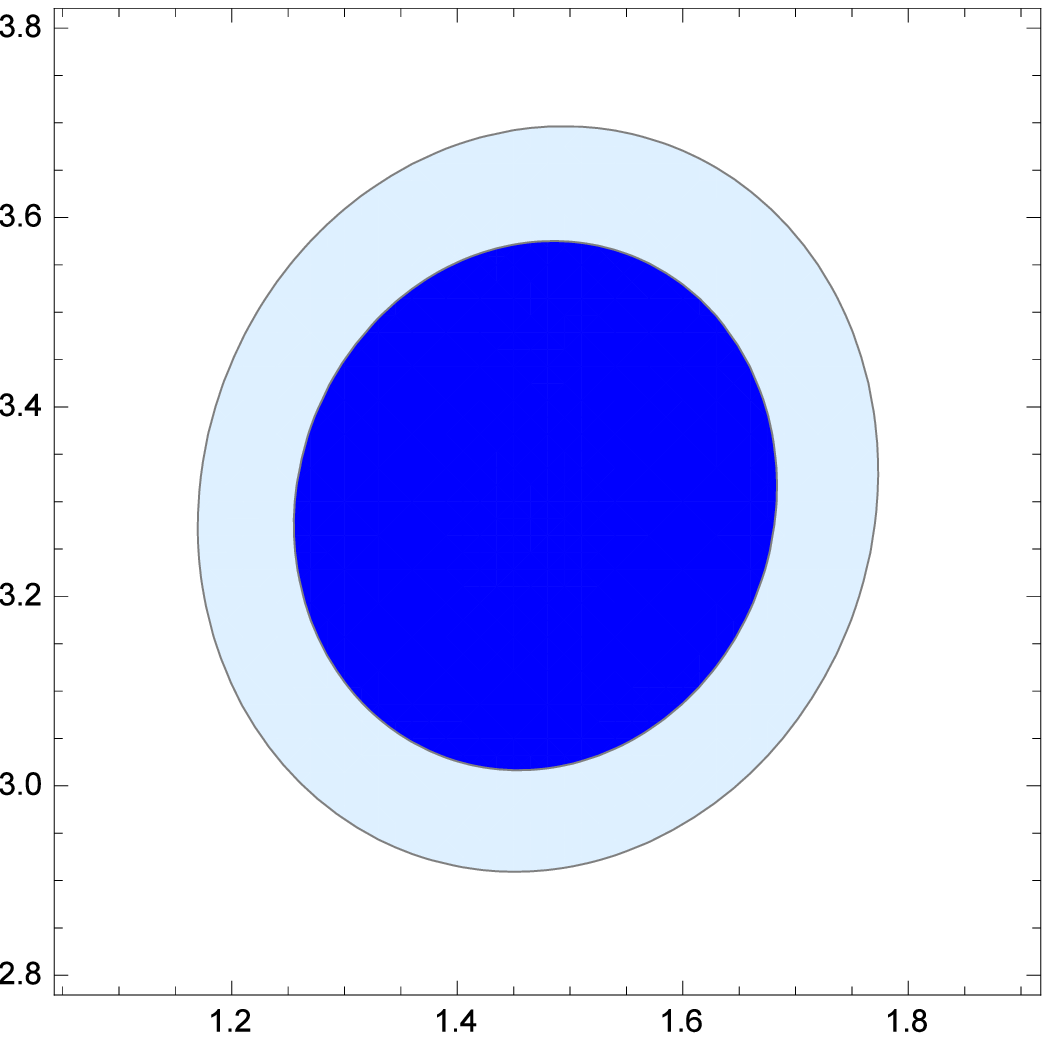}\hspace{.00001cm}
\includegraphics[width=.125\textwidth]{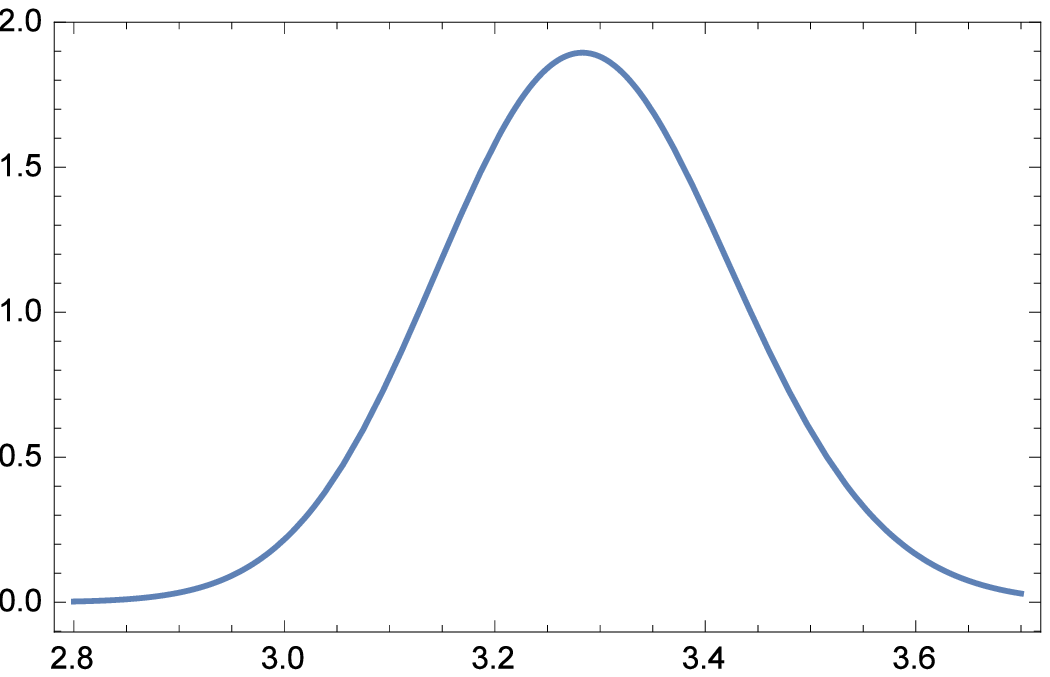}\hspace{.00001cm}~~~~~~~~~~~~~~~~~~~~~~~~~~~~~~~~~~~~~~~~~~~~~~~\\
 \includegraphics[width=.143\textwidth]{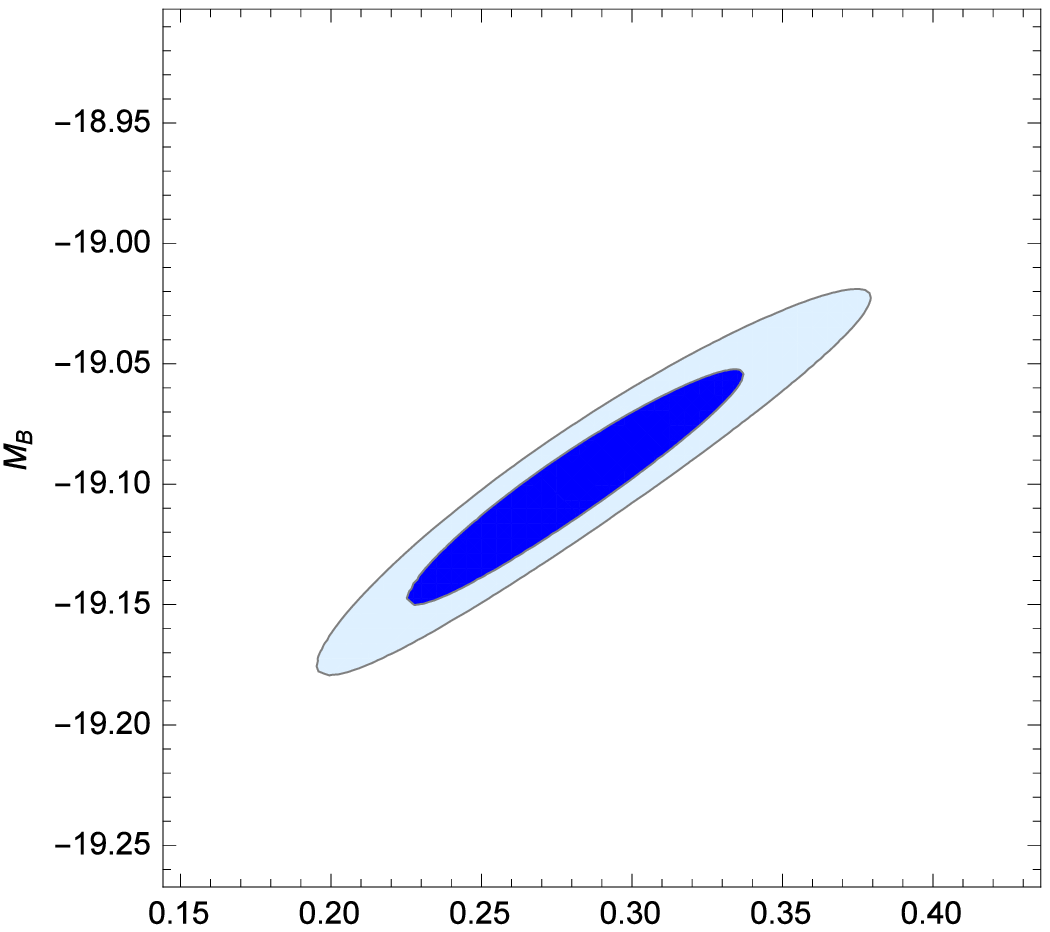}\hspace{.000001cm}
\includegraphics[width=.13\textwidth]{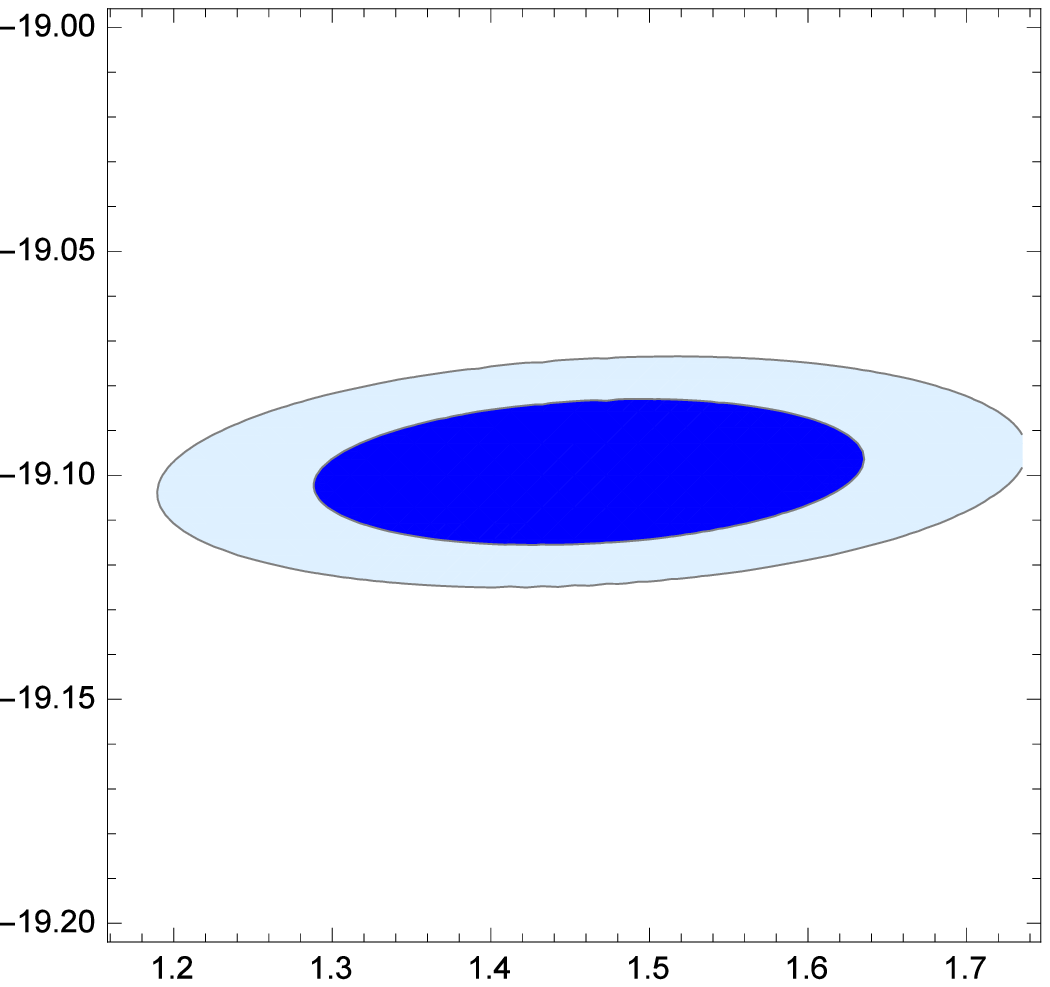}\hspace{.000001cm}
\includegraphics[width=.13\textwidth]{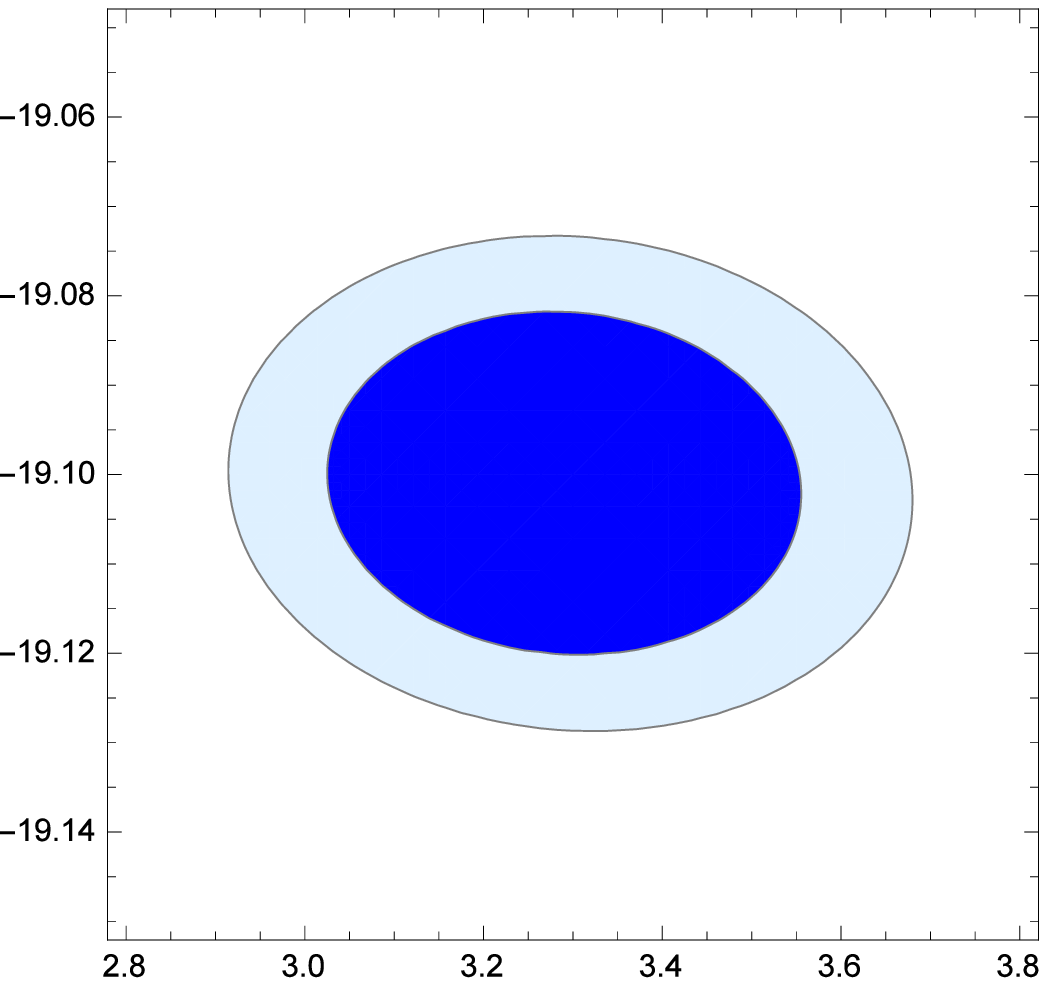}\hspace{.000001cm}
\includegraphics[width=.125\textwidth]{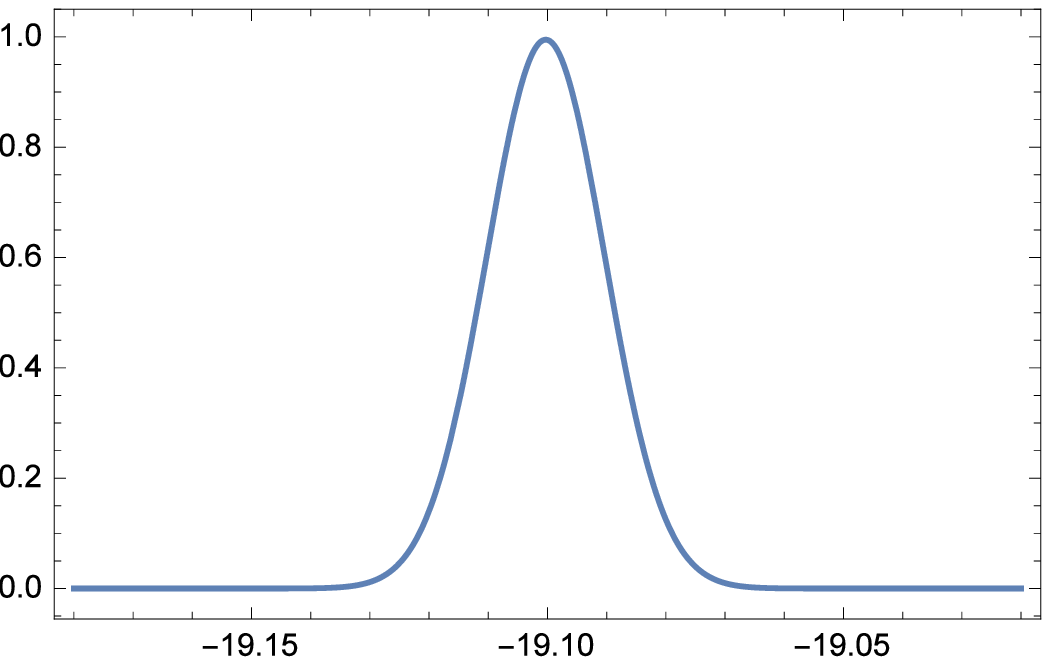}\hspace{.000001cm}~~~~~~~~~~~~~~~~~~~~~~~~\\
\includegraphics[width=.14\textwidth]{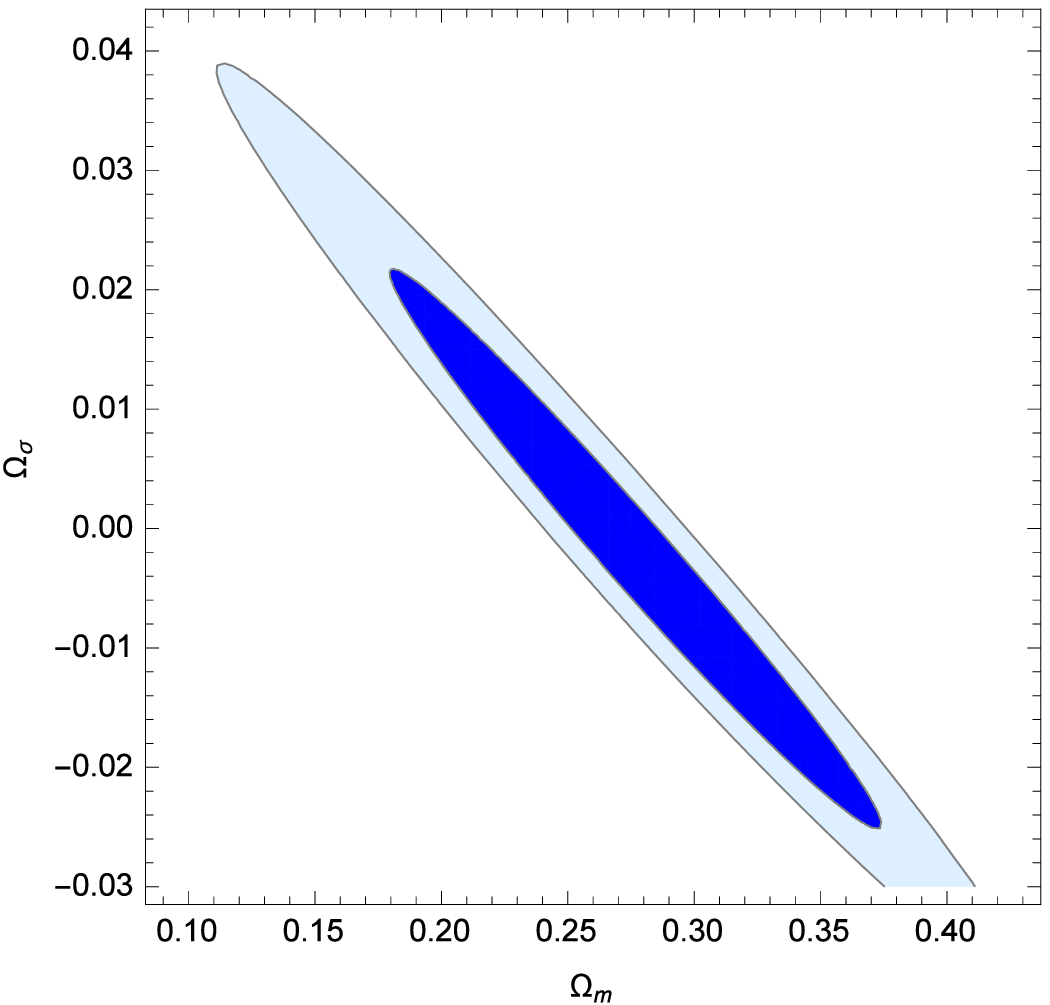}\hspace{.00001cm}
 \includegraphics[width=.13\textwidth]{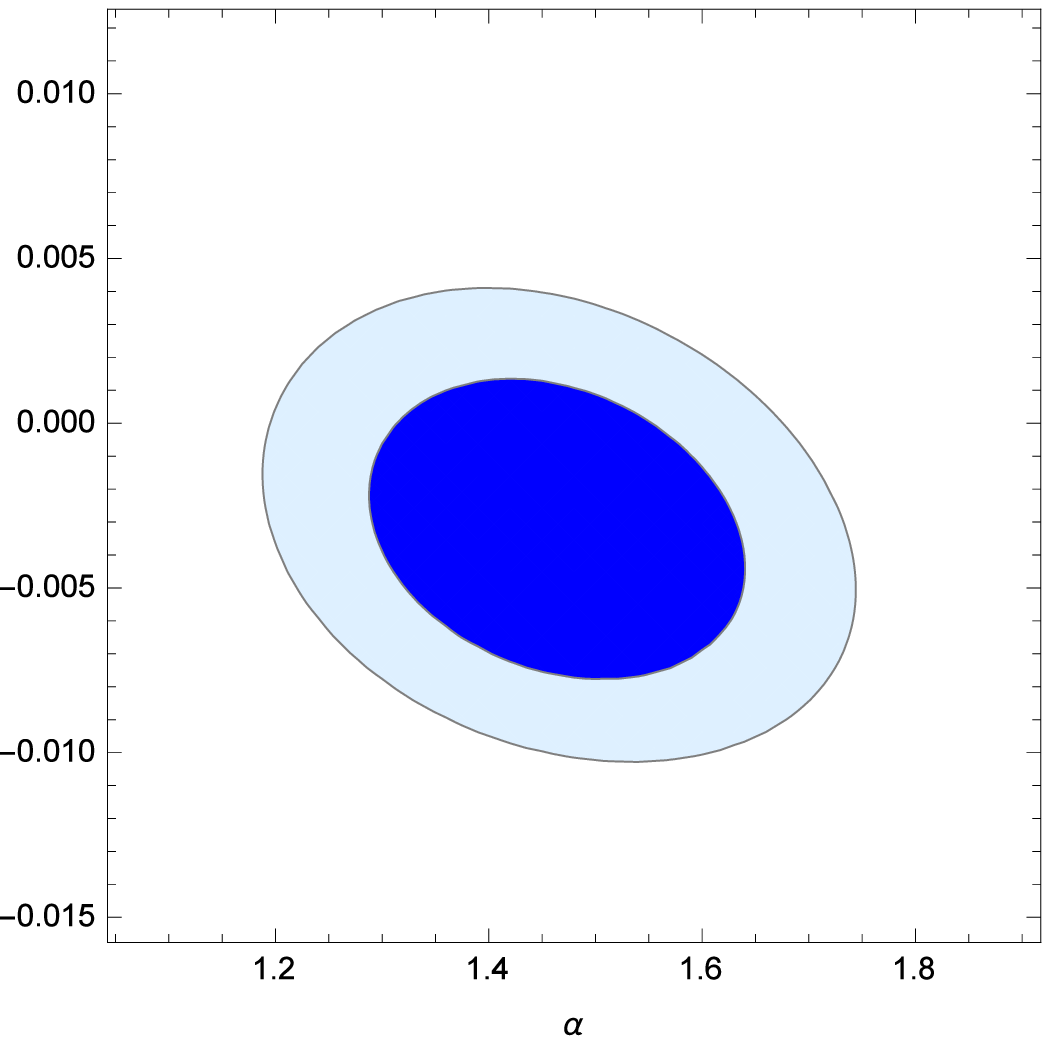}\hspace{.00001cm}
\includegraphics[width=.13\textwidth]{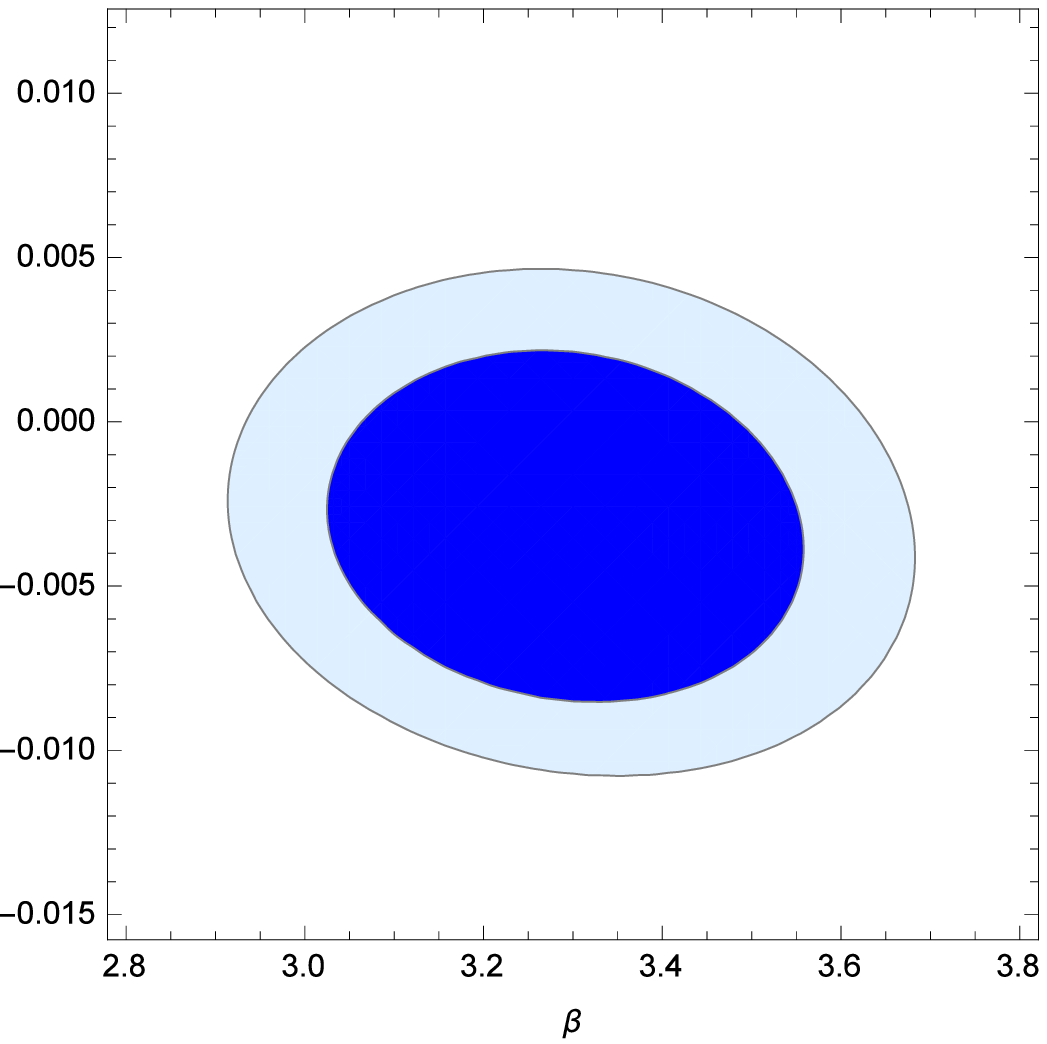}\hspace{.00001cm}
\includegraphics[width=.13\textwidth]{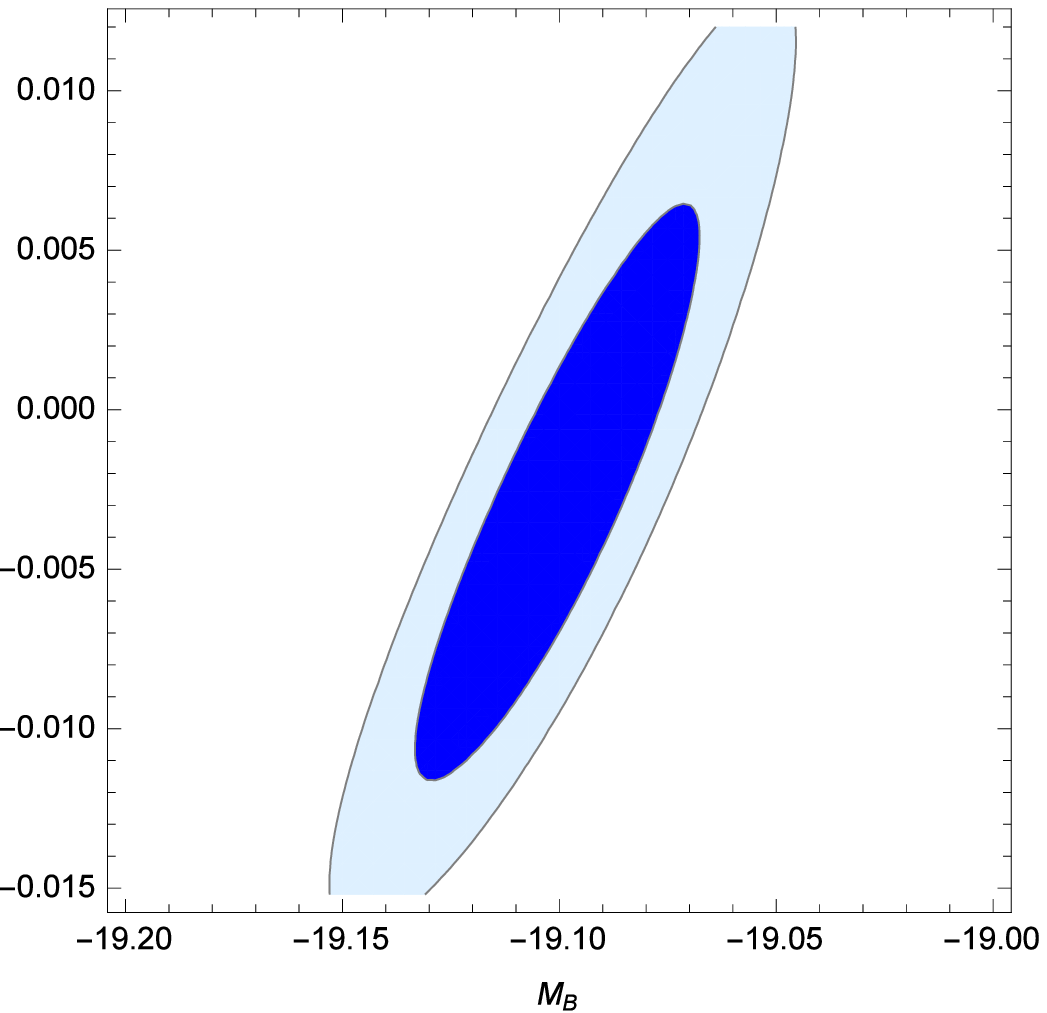}\hspace{.00001cm}
 \includegraphics[width=.14\textwidth]{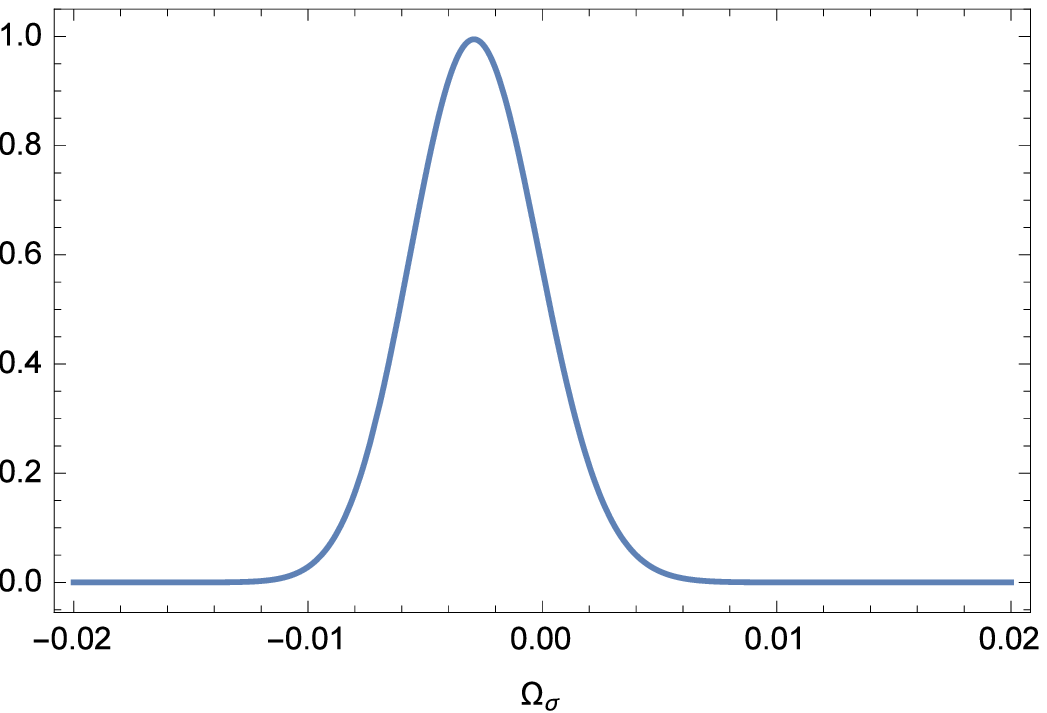}\hspace{.00001cm}
\caption{Normalized likelihood distributions and 2D joint distributions with 68\% and 95\% confidence contours for the $\Lambda$CDM fit parameters, using the simulated sample with JLA sample of 238 SNe Ia. }
 \label{fig:7}
 \end{figure}
\begin{widetext}
\begin{center}
\begin{table}
{ \bf Table III:} Best Fits for Different Cosmological Models  for JLA SNe Ia.  \\
\begin{tabular}{cccccccc}
\hline\hline  {\bf Model}        &{\bf $\Omega_{\sigma_0}$} & {\bf $\Omega_{m_0}$ }    &  {\bf $\alpha$}& $\beta$ &$M_B$&  $\chi_{min}^2/dof$ &{\bf Other Best Fit Parameters} \\
\hline $\Lambda$CDM& $-0.003\pm0.033$   &$0.28\pm 0.202$ &$1.4022\pm0.111$      &$3.279\pm0.145$  &$-19.1\pm0.067$    &$0.99$ &$-$ \\
\hline
 $w$CDM &$-0.0027\pm0.032$   &$0.279\pm 0.466$ &$1.4\pm0.109$      &$3.175\pm0.147$  &$-18.9\pm0.054$    &$0.997$ &$w=-0.9999\pm0.866$ \\
 \hline
 q-$\Lambda$ &  $0.006\pm0.045$   &$0.252\pm 0.318$ &$1.4003\pm0.11$       &$3.1751\pm0.151$  &$-19.104\pm0.105$    &$1.001$ &$a_1=(-1.6\pm2.2)\times 10^{-4}$ \\&&&&&&& $b_1=3.101\pm0.02$ \\
 \hline
 Linear  &$0.004\pm0.06$   &$0.39\pm0.126$ &$1.169\pm0.109$       &$3.14\pm0.132$  &$-20.086\pm0.073$    &$1.022$ &$w_0=-0.788\pm0.11$ \\&&&&&&& $ w_a=0.307\pm0.39$ \\
 \hline
 GCG & $0.0014\pm0.039$   &$0.278\pm 0.323$ &$1.192\pm0.142$       &$3.043\pm0.178$  &$-19.493\pm0.064$    &$1.227$ &$A=0.968\pm0.005$ \\&&&&&&& $\alpha=18.5\pm2.378$ \\
 \hline
 CG & $0.075\pm0.032$   &$0.34\pm 0.051$ &$1.238\pm0.13$       &$3.129\pm0.186$  &$-19.508\pm0.042$    &$1.071$ &$A=0.974\pm0.084$ \\
\hline\hline
\end{tabular}
\end{table}
\vspace{-0.2cm}
\end{center}
\end{widetext}

 \section{Conclusion}
In this work, we have investigated the  anisotropy effects on the various cosmological models, by using the   two
 SNIa datasets available in the work: the SNIa data by Tonry \textit{et al}. and Barris \textit{et al}.   in the
redshift range $0\leq z\leq1.75$ and the most recent SNLS dataset (238 data points  $0.15< z<1.1$). We  have   fitted these models using the MLE  and compare our results with the inference made using luminosity distances measured with both SNIa     data points.  We have used representative
DE parametrizations to examine the consistency among the two datasets in constraining the corresponding parameter values.
Based on these properties, we have evaluated the quality of fit of several $H(z)$  DE models
and it was compared the observed expansion rate $H(z)$ with that predicted by the various models proposed in Table   I.
  In comparing the quality of fit we have used the value of   $\chi_{min}^2$ and the p-test in BI models.  We have  reported the model independent reconstruction of the cosmic EoS and deceleration parameter of DE. Our main conclusions can be summarized as follows. \\
\begin{itemize}
\item
The first and simplest studied model was SCDM model  and we found $\chi^2_{min}=468.994$ for it, by using  the data from 194 SNIa \cite{47}. Fig. 1 exhibits that applying anisotropy for the data fitting is not adequate for SCDM model.

\item
Another simple model was the $\Lambda$CDM model in which $\Omega_{\sigma_0}$ parameter is the free parameter. The best is equal to $\Omega_{\sigma_0}=0.0128$ with $\chi^2_{min}=197.559$ and $\Omega_{m_0}$. We also showed that applying anisotropy on this model leads to the  good fit using the SNIa data (see Fig. 1).
\item
We found the best fitted parameter in $w$CDM model by fixing $\Omega_{\sigma_0}=0.013$ is $w=-1.0146\pm 0.0806$ with $\chi^2_{min}=197.276$. This result shows that the minimization of the $w$CDM model improves the data fitting in $w$CDM model in comparison with the $\Lambda$CDM model. In other words, EoS parameter can cross the phantom divide  line, if the   $w$CDM model is fitted using the data obtained from SNIa, and we found $q_0=-0.555$ and $z_t=0.449$ at the first level of error.
\item
In  the $\Lambda$-q model the best fitting parameter, for SNIa data and by considering anisotropy was     $a_1=2.3^{+1.3}_{-0.1}\times 10^{-4}$ and $b_1=3.1^{+0.421}_{-0.893}$ with $\chi^2_{min}=196.967$. In this model, the EoS parameter  cannot cross the phantom divide  line, i.e. this model is in an agreement with the
quintessence model. As Fig. 2 suggests, applying anisotropy leads to a faster accelerated expansion for the universe. We also found that the amount of $q$ at the current  time is $-0.51$ and the transition redshift is $z_t=0.404^{+0.021}_{-0.023}(1\sigma')^{+0.046}_{-0.042}(2\sigma')$.
\item
We obtained $w_0=-1.261^{+0.003}_{-0.024}$ and $w_a=1.417^{+0.147}_{-0.021}$ with $\chi^2_{min}=196.58$, by applying anisotropy to the linear model. Table I shows a better fitting in comparison to other two models. In this table the minimum amount of  $\chi^2$ has been presented. Therefore, among the DE models studied in present work,  we are interested in the linear model by using SNIa data. The EoS parameter can also cross   phantom divide  line in this model. We calculated the $q$ parameter and  the transition redshift $z_t$ in this model and its results are $q_0=-0.781$ and $z_t=0.315$.

\item
We have analyzed the currently available 194 supernova data points within the framework of the generalized Chaplygin gas and  Chaplygin gas models. We have considered both, flat and non-isotropic cases, and   found the best fit parameters in the generalized Chaplygin gas model are $\alpha=15.1^{+0.121}_{-0.092}$ and $A=0.999777^{+0.000223} _{-0.000448}$   with $\chi^2_{min}=196.791$. The generalized Chaplygin gas model tends to be the  $\Lambda$CDM model in order to recover the standard cosmology at early times.    In other words,  our results show that the generalized Chaplygin gas model is almost the same as the $\Lambda$CDM model than the   Chaplygin gas  model.   It is easy to see that the best fit $w (z)$ can not cross $-1$ as it evolves with the redshift $z$, and the present best fit value $w(0)\sim -0.999$ as shown in Fig. 4a. The transition redshift when the universe underwent the transition from deceleration to acceleration is found to be $z_t=0.4$. We also find that the present deceleration parameter is $q_0=0.511$ (see Fig. 4b). But in the case of  the Chaplygin gas model the best-fit values of parameter  is  $A = 0.999\pm0.0225$ with $\chi^2_{min}=197.565$ at the $1\sigma$ level.
In Table II we present values of the EoS  and deceleration parameters  for    generalized Chaplygin gas and   Chaplygin gas model.
 It can be seen clearly that  the generalized Chaplygin gas model is preferred  of the  Chaplygin gas model by SNIa data. The fact that the properties of the Chaplygin gas interpolate between those of SCDM and a $\Lambda$CDM led to the hope that the Chaplygin gas might provide a
conceptual framework for a unified model of DM and DE.
 \item
 We also consider the   light curve fitters  for our analysis.  It is intended to estimate three parameters (magnitude, shape and colour) that can be subsequently linearly combined to determine the luminosity distances of the SNe Ia.    Unlike  previous section, we have assumed prior corresponding to  $\Omega_{m_0}=0.3$,   this approach can thus adversely affect the validity of the fitting method and lead to compromised or misleading results. Therefore to counteract this problem,   we  does not require the prior on  all parameters, fit by the SNLS dataset.   Consider no prior of $\Omega_{m_0}$ and the SN light-curve model,  we re-optimize parameters by carrying out an MLE in any situation where the parameters include an
unknown intrinsic dispersion. The commonly used method, which estimates the dispersion by requiring the reduced $\chi^2$ to
equal unity, does not take into account all possible covariances among the parameters. We have found that, when the parameter
optimization is handled via the joint likelihood function, DE  models  fit their individually optimized data very
well. In the $\Lambda$CDM model, the JLA sample provides a measurement of the reduced matter density parameter $0.28\pm 0.202$, which is in good agreement with the recent measurement from the JLA and lowz+SNLS \cite{86},  SN-stat \cite{87} and SNeIa+CMBShift \cite{88}.
\end{itemize}
Finally, seven models and considering anisotropy on them were investigated.  By comparing between Tables I and III, it was concluded that fitting the
SNIa data and applying anisotropy, leads to an improvement of the SNIa data (except for the
SCDM model).  The best fit  parameters in a $\Lambda$CDM parametrization without a $\Omega_{m_0}$ prior show interesting differences between the Tonry and Barris  \textit{et al}.   and the JLA SNIa datasets. Thus, our study shows that the anisotropy effects on the $\Lambda$CDM DE model is a very
good fit to the latest JLA SNIa data.


\end{document}